\documentclass[aps,superscriptaddress,showpacs,amsmath,amssymb,amsfonts,altaffillsymbol,twocolumn,notitlepage,prl]{revtex4-1}

\usepackage{graphicx}
\usepackage{float}
\usepackage{dcolumn}
\usepackage{bm}
\usepackage{color}
\usepackage{soul}
\usepackage{xcolor,hyperref}

\graphicspath{{./figures/}}

\begin{document}

\title{Self-organization and memory \\ in a cyclically driven elasto-plastic model of an amorphous solid}

\author{D. Kumar}%
\email[Corresponding author: ]{dheeraj.kumar@espci.fr}
\affiliation{PMMH, CNRS, ESPCI Paris, Universit\'e PSL, Sorbonne Universit\'e, Universit\'e Paris Cit\'e, France}%

\author{Muhittin Mungan
}
\affiliation{Institute for Biological Physics, University of Cologne, Z{\"u}lpicher Stra{\ss}e 77, K{\"o}ln, Germany}

\author{S. Patinet}%
\affiliation{PMMH, CNRS, ESPCI Paris, Universit\'e PSL, Sorbonne Universit\'e, Universit\'e Paris Cit\'e, France}%

\author{M. Mert Terzi}%
\affiliation{PMMH, CNRS, ESPCI Paris, Universit\'e PSL, Sorbonne Universit\'e, Universit\'e Paris Cit\'e, France}%

\author{D. Vandembroucq}%
\email[Corresponding author: ]{damien.vandembroucq@espci.fr}
\affiliation{PMMH, CNRS, ESPCI Paris, Universit\'e PSL, Sorbonne Universit\'e, Universit\'e Paris Cit\'e, France}%


\begin{abstract}
The mechanical behavior of disordered materials such as dense suspensions, glasses or granular materials depends on their thermal and mechanical past. 
Here we report the memory behavior of a quenched mesoscopic elasto-plastic (QMEP) model. After prior oscillatory training, a simple read-out protocol gives access to both the training protocol's amplitude and the last shear direction. 
The memory of direction emerges from the development of a mechanical polarization during training. 
The analysis of sample-to-sample fluctuations gives direct access to the irreversibility transition. 
Despite the quadrupolar nature of the elastic interactions in amorphous solids, a behavior close to Return Point Memory (RPM) is observed. The quasi RPM property is used to build a simple Preisach-like model of directional memory.
\end{abstract}


\maketitle

From toothpaste and polymers to glass and concrete, most everyday life materials are disordered: they do not have a regular structure at the atomic or particle scale. 
Unlike crystalline materials, for which there is a single, well-defined thermodynamic state, disordered materials are out of equilibrium and can be found in a myriad of different states. Not all these states are equivalent so that the physical and mechanical properties of disordered solids or complex fluids usually depend on how they were prepared and processed. In other words, most of these materials carry at least a partial memory of their thermal and mechanical past.

However, what exactly is the nature of this memory? What information can be recorded and subsequently read out in a disordered material?
How is it connected to the underlying structure of the material? Over 
the last decade, the characterization and understanding of 
mechanical memory in materials has motivated a growing number of
studies, as recently reviewed in ~\cite{Keim-RMP19,Paulsen-Keim-ARCM24}.
In particular, experimental works on colloidal suspensions~\cite{pine2005, corte2008random,  keim-RSC13, Keim-Arratia-PRL14,Paulsen-Keim-Nagel-PRL14, mukherji2019strength,keim2020global}, as well as particle simulations of disordered solids upon oscillatory driving~\cite{regev2013onset,Fiocco-PRE13,Priezjev-PRE16}, demonstrated the existence of a dynamic transition between a reversible state at low amplitude,  in which the same sequence of plastic events repeats periodically and a diffusive state at high amplitude. 
In a parallel line of research, experimental and numerical studies explored the limit cycles reached after oscillatory driving in order to understand in more detail the memory capacity of disordered solids \cite{keim2011generic, Fiocco-PRL14, adhikari2018memory, regev2021topology, benson2021memory, keim2020global, arceri2021marginal, keim2022ringdown, shohat2023dissipation}. 
This research demonstrated that, when combined with a properly defined read-out protocol, it is possible to record and subsequently retrieve the amplitude of oscillatory shear that had been applied beforehand, {\it i.e.} during ``training", to the material. 
More specifically, after training, numerical or experimental samples are subject to a sequential read-out protocol consisting of a sequence of oscillations with growing amplitude. 
After each reading cycle, the difference between the trained and current states are compared by a suitably defined distance. The evolution of this distance with the read-out cycle shows a well-defined minimum when the read-out amplitude matches the training amplitude.

Memory in periodically driven physical systems has been traditionally associated with hysteretic behavior in magnets, leading to return-point memory (RPM) \cite{preisach1935magnetische,everett1954general,barker1983magnetic, sethna1993hysteresis}. During the mechanical deformation of disordered materials, hysteresis results from local
instabilities, such as the buckling of creases of a crumpled thin elastic sheet~\cite{shohat2022memory,Mungan-PNAS22} or the local plastic rearrangements of the amorphous structure for glasses~\cite{Falk-Langer-PRE98,Barbot-PRE18}. Following the spirit of the Preisach model that was initially developed to describe hysteresis in magnets \cite{preisach1935magnetische,everett1954general,barker1983magnetic,bertotti2006book, brokate2012hysteresis}, 
mechanical memory has often been modeled as a collection of {\it hysterons}, i.e. elementary units of hysteresis \cite{keim2020global,terzi2020state,bense2021complex, keim2021multiperiodic, lindeman2021multiple, van2021profusion, szulc2022cooperative, muhaxheri2024bifurcations}.
However, unlike spin systems, plastic events in disordered materials are coupled through an elastic interaction which is long-range, anisotropic, and is not expected to admit a no-passing (NP) property \cite{middleton1992asymptotic}. The NP property asserts the existence of a partial ordering of states preserved under the driving and is a sufficient condition for the emergence of RPM \cite{sethna1993hysteresis, munganterzi2018}. An
immediate consequence for disordered materials is that, due to the absence of NP, one would a priori not expect RPM to emerge. Nevertheless, approximate RPM has been observed in such systems -- both experimentally and numerically \cite{keim2020global,mungan2019networks}.

Models of collections of interacting hysterons or Ising-like spins, as considered, for example, in \cite{lindeman2023isolating, liu2024controlled, sirote2024emergent}, may then be too
simple to cope with the complex mechanical memory of disordered
materials. To overcome these limitations, an appealing alternative
is to resort to mesoscopic elasto-plastic models~\cite{Nicolas-RMP18}. These lattice models rely on local threshold dynamics in which each cell can experience a local slip whenever the local stress reaches a prescribed threshold value. Each cell rearrangement is, in addition, coupled with a quadrupolar long-range redistribution of elastic stresses via an Eshelby-type kernel~\cite{Eshelby-57,Picard-EPJE04,Tyukodi-PRL18}. 
Despite their simplicity, upon monotonic loading, these mesoscale models reproduce most of the phenomenology of the plastic behavior of amorphous plasticity, such as avalanches~\cite{Wyart-PNAS14}, localization~\cite{VR-PRB11} and creep~\cite{Castellanos-PRL18}.
It was recently shown that the reversible plastic behavior observed upon cyclic loading and the associated irreversibility transition could
also be reproduced with such type of models \cite{Khirallah-Maloney-PRL21, Liu-Ferrero-JCP22, Kumar-JCP22}. 
Among its other
advantages, the discrete nature of mesoscopic models makes them a natural tool to
build the transition graphs recently introduced to study and
characterize the complex landscape of amorphous
solids~\cite{mungan2019networks,regev2021topology,Kumar-JCP22}. 

In the following, we give a short description of the {\it quenched mesoscopic elastoplastic model} (QMEP), which we introduced recently, allowing us to study the mechanical memory behaviour under oscillatory shear~\cite{Kumar-JCP22}. After defining a proper read-out protocol, we show that  the amplitude and direction of past oscillatory driving can be retrieved. Next, we analyze sample-to-sample fluctuations under read-outs, and demonstrate how these fluctuations are connected with the irreversibility transition. In the subsequent sections we analyze the mechanical properties of the solid and how these lead to  memory formation. We first focus on the annealing of local stresses under oscillatory shear and link this to the observed RPM-like behavior. We then formulate a Preisach-like model of directional memory whose predictions agree well with our numerical results.  We conclude by outlining perspectives that emerge from our work.

\section*{Quenched Mesoscopic Elasto-Platic models}

The quenched mesoscopic elasto-plastic model (QMEP) of  Ref.~\cite{Kumar-JCP22} follows the spirit of early depinning-like models of amorphous plasticity~\cite{BVR-PRL02}.
The plastic deformation of dense amorphous solids results from a series of localized rearrangements of the disordered structure. These local plastic events are the building blocks of the mesoscopic elasto-plastic models. Discretization is performed on a lattice at a mesoscopic length scale, which is large enough to experience a plastic event but still small enough so that only one such event can occur at a time~\cite{Barbot-PRE18,Patinet-CRPhys21}. The physics of the model relies on the coupling between local threshold dynamics and an elastic interaction. The latter leads to a redistribution of local stresses whose general features derive directly from the solution of the eigenstrain Eshelby inclusion problem~\cite{Eshelby-57}. Part of the complex phenomenology exhibited by EP models is due to the long-range and quadrupolar character of the stress redistribution: depending on the distance and orientation, a plastic event tends to either facilitate or prevent plastic events at distant cells~\cite{Wyart-PNAS14,TPRV-PRE16,Tyukodi-PRL18,Martens-PRL24}.
The mesoscopic nature of the model here arises from the fact that the disordered thresholds assigned to each cell emerge from the details of the amorphous structure on a microscopic scale~\cite{PVF-PRL16,Barbot-PRE18}, while the interaction between the cells is obtained by calculating elasticity on a continuous scale.

Although noise and disorder have been essential ingredients of both mesoscopic elasto-plastic models~\cite{BulatovArgon94a,BVR-PRL02,Picard-PRE02} as well as rheological mean-field models~\cite{Hebraud-Lequeux,Sollich-PRL97}, implementations and interpretations have been so diverse that, to this day, no clear consensus has emerged on their role and importance in the various features of amorphous plasticity. In this context, a key and entirely original feature of the model introduced in Ref.~\cite{Kumar-JCP22} is the quenched character of the plastic threshold disorder.

\begin{figure}[t!]
\centering
\includegraphics[width=0.85\columnwidth]{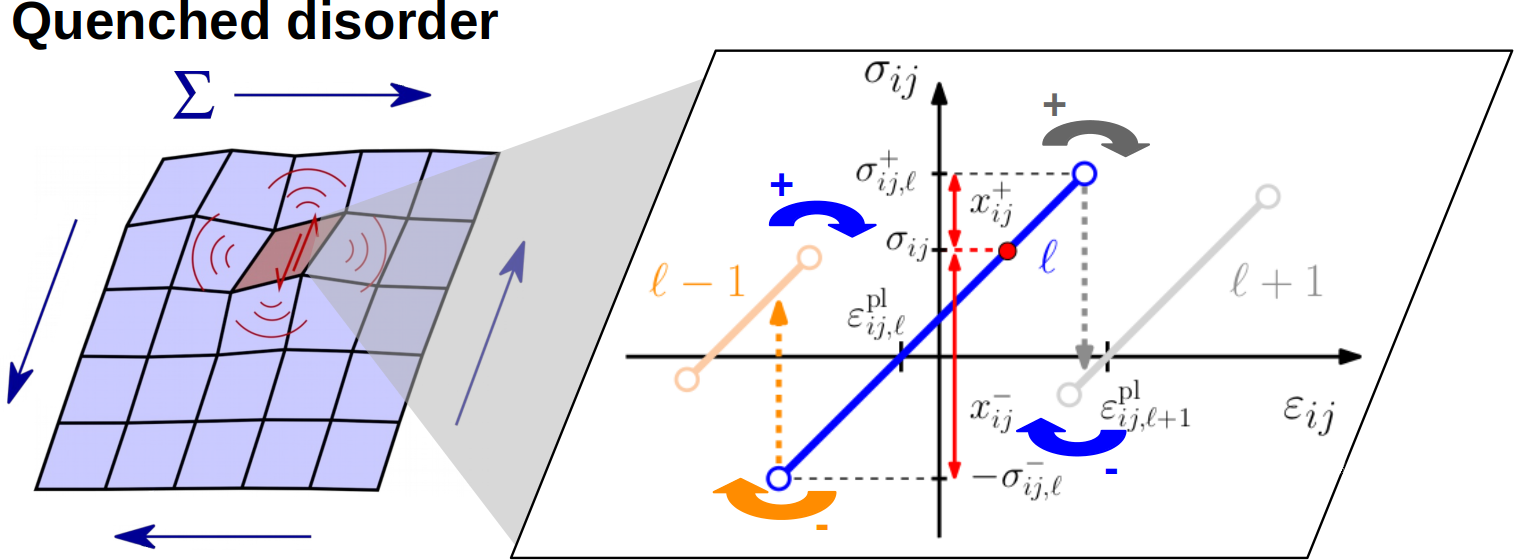}
\caption{{\bf Mesoscale modeling of an amorphous solid.} (Left) Lattice of cells. (Right) The response of each cell $(i,j)$ to an external stress $\Sigma$ consists of a sequence of elastic branches, labeled $\ldots, \ell-1, \ell$ and $\ell + 1, \ldots$, which are limited by local stress thresholds in the forward $\sigma_{ij}^+$ and reverse $\sigma_{ij}^-$ directions of shear. The collection of all local elastic branches across cells constitutes the quenched disordered landscape of the model.} 
\label{quenched-disorder}
\end{figure}

In the QMEP model we consider a random distribution of plastic thresholds for 
 a two-dimensional amorphous solid that has been spatially 
coarse-grained into an $N \times N$  lattice $\Lambda$ of mesoscale 
cells indexed as $(i,j)$, as illustrated in Fig.~\ref{quenched-disorder}. 
To each cell $(i,j)$, we assign a pair $(\sigma_{ij}^-,\sigma_{ij}^+)$ of plastic thresholds which bound the range of mechanical stability of the cell and constitute a {\em local elastic branch}. If
upon driving in the {\em forward (reverse)} direction, the local stress
$\sigma_{ij}$ reaches the upper (lower) bound $\sigma_{ij}^+$
($\sigma_{ij}^-)$ of the elastic branch, the system jumps toward a
new local elastic branch. This branch in turn is bounded by its own pair of plastic thresholds, which are
drawn from the same random distribution of thresholds as the original one.  

In particular, we consider a quenched disorder~\cite{Kumar-JCP22}, as illustrated in Fig.~\ref{quenched-disorder}, so that the stack of local elastic branches, labeled as $\ldots, \ell-1, \ell, \ell + 1, \ldots$, and hence their properties, are {\em frozen}. We call this the (local) stress landscape of a given cell. If after a forward drive, causing the cell to be driven from elastic branch $\ell$ to $\ell + 1$, the system is subject to a reversal of driving direction, it must first revisit the local elastic branch $\ell$ with its corresponding stress thresholds. Thus, plastic events may be reversible.
Such reversible local plastic events were observed in early molecular dynamics simulations and experiments by Lundberg {\it et al.}~\cite{OHern}.  In order to characterize the stress landscape seen by the individual cells $(i,j)$, we label their local elastic branches as $\ell_{ij}$ so that the local branch configuration of all cells of the lattice is given by the collection of integers $(\ell_{ij})_{(i,j) \in \Lambda}$.
We can think of the triplet $(i,j, \ell_{ij})$ as a 3D embedding of the 2D lattice $\Lambda$ of mesoscopic cells~\cite{TPRV-PRE16,Wyart-PNAS14}. This triple, in turn, establishes the local elastic branches via the parameters such as $(\sigma_{ij,\ell_{ij}}^-,\sigma_{ij,\ell_{ij}}^+)$. Thus, regarding the 3D embedding, the latter are quenched random variables. When projected down to the 2D lattice  of cells $(i,j)$, the local branch index $\ell_{ij}$ acts like the height function of an elastic interface evolving in a {\it quenched} disordered landscape (see Ref.~\cite{Lerbinger-CRPhys23} for a
discussion). 


The initial condition, {i.e.} the plastic thresholds $\sigma_{ij}^\pm$ and local stress fields $\sigma_{ij}$ obtained after a randomizing sequence of effective thermal and aging steps leads to a local branch configuration that we define as $\ell_{ij} \equiv 0$ for all cells $(i,j)$. By varying the thermalization and aging protocol, we can tune the systems' effective age and mechanical behavior from the poorly aged and very ductile case to the very aged and highly brittle case. In the following, we restrict our study to poorly aged (PA) samples, whose preparation protocol has been described in Ref.~\cite{Kumar-JCP22,Kumar-PhD23}.
 
We simulated systems of size $16\times 16$, $32\times 32$ and $64\times 64$, and used a Weibull distribution for the thresholds: $P(\sigma^\pm) = 1 -e^{(\sigma^\pm/\lambda)^\kappa}$ with $\kappa=2$ and $\lambda=0.1$. Note that the stress values are rescaled with the shear modulus $\mu$. The plastic strain increment $\Delta \varepsilon = \varepsilon^{pl}_{ij,\ell+1} - \varepsilon^{pl}_{ij,\ell}$ between two neighbor elastic branches $\ell$ and $\ell+1$ is also a random variable. We choose it to be correlated to the two stress thresholds associated with the transition $\ell \to \ell+1$: $\Delta \varepsilon$ is drawn from a uniform distribution in $[0,\Delta \varepsilon_{max}]$ with $\Delta \varepsilon_{max}=\eta (\sigma^+_{ij,\ell} + \sigma^-_{ij,\ell+1})/2$. The tunable parameter $\eta$ controls the strength of the elastic interaction, and we set $\eta=1$~\cite{Kumar-JCP22,Kumar-PhD23}. 

\section*{Memory of amplitude of past oscillatory shearing}

In order to probe the memory behavior, the systems are first subjected to a {\it training protocol}, followed by a {\it read-out protocol}. 
Here, we discuss the reversible plastic behavior that can be reached upon training by application of cyclic shear; we then present two different read-out protocols and discuss their performance in retrieving the amplitude of the training protocol. We finally discuss the fluctuation behavior of the read-out response and link it with the irreversibility transition.

\subsection*{Oscillatory training: reversible plasticity and irreversibility transition}
The training protocol consists of a sequence of $\mathcal{N}_T$ shear cycles of oscillatory shear $0 \to \varepsilon_T \to - \varepsilon_T \to 0$,  where $\varepsilon_T$ is the amplitude.
In Ref.~\cite{Kumar-JCP22}, we investigated the evolution of the QMEP model towards cyclic response under oscillatory shear. We found that the system quickly locks into a limit cycle at low training amplitudes, and the response is perfectly elastic. This behavior continues up to a shear amplitude $\varepsilon_{hys}$, where a hysteresis loop opens up, and plasticity emerges, albeit in a {\it reversible} way ~\cite{lundberg2008reversible, Keim-Arratia-PRL14,Khirallah-Maloney-PRL21,EVM-preprint22}: 
as the limit-cycle is traversed, the sequence of plastic events takes place in such a way as to precisely compensate each other so that at the end of the limit-cycle, the system has returned to the same configuration as at its beginning. This means that the same sequence of plastic events occurs from cycle to cycle, hence the term {\it reversible plasticity} ~\cite{lundberg2008reversible}. Beyond the hysteresis transition $\varepsilon_{hys}$, the number of training cycles needed to reach reversibility gets larger with increasing amplitude $\varepsilon_T$~\cite{EVM-preprint22}. At the same time,  the eventually attained limit-cycles involve an increasingly larger number of plastic events, while the periodicity of the cyclic response starts to span multiple driving cycles -- a phenomenon called multi-periodicity or subharmonicity ~\cite{deutsch2003subharmonics, Lavrentovich2017multiperiodic}. We will use {\em monoperiodic} to indicate that the cyclic response period coincides with that of the driving.

This behavior continues up to a strain amplitude $\varepsilon_{irr}$. It marks the onset of the {\it irreversibility transition}  beyond which no limit cycle can be found, and diffusion starts to take place~\cite{Fiocco-PRE13,regev2013onset,Kawasaki-Berthier-PRE16}. 
In Ref.~\cite{Kumar-JCP22}, we defined $\varepsilon_{irr}$  in a statistical way. Given a maximum number $\mathcal{N}_T$ of driving cycles, $\varepsilon_{irr}$ is the training strain amplitude at which $50\%$ of the realizations lead to a cyclic response. The amplitude $\varepsilon_{hys}$, marking the transition from purely elastic to plastic cyclic response, is defined similarly. 

While we present only results obtained from poorly-aged (PA) realizations of size $16\times 16$, we would like to note that qualitatively similar responses were also obtained for PA systems of sizes $32\times 32$ and $64\times 64$. For the $16\times 16$ PA glasses we chose $\mathcal{N}_T=10^4$ and find that $\varepsilon_{hys}=0.025$ and $\varepsilon_{irr}=0.0689$. Unless otherwise noted, all results to be shown are averages obtained from $1000$ realizations of trained glasses.


\subsection*{Read-out protocols: Implementation and results}
The trained samples are then subjected to a {\it read-out protocol} in which we apply a single cycle of oscillatory shear at {\em read-out amplitude} $\varepsilon_R$. As described below, we then use a quantitative observable to characterize the proximity of the states of our glass prior to and after the read-out, labeling these as $T$ and $R$, respectively. Experimentally, a {\it sequential} read-out protocol is typically used~\cite{Paulsen-Keim-Nagel-PRL14, Keim-RMP19,Paulsen-Keim-ARCM24}. This consists of applying to the trained state $T$ a sequence of shear cycles of increasing amplitude $\varepsilon_R$. At the end of each cycle, the system's state is compared with $T$. Numerically, it is also possible to resort to a {\it parallel} read-out protocol, which consists of taking multiple replicas of the trained state $T$ and applying to each a single cycle of oscillatory shear with some amplitude $\varepsilon_R$~\cite{Fiocco-PRL14}.  
The state $R$ that each replica attains at the end of its read-out cycles is then compared with the trained state. Empirically, for both training protocols, maximum proximity between trained and read-out configurations is found to occur when $\varepsilon_R \approx \varepsilon_T$. In this sense, the trained systems have a memory of $\varepsilon_T$, which subsequently can be read out \cite{Keim-RMP19}.

In particle simulations and experiments on colloidal suspensions, the proximity between two configurations is quantified by the particles' mean
square displacement (MSD), which involves comparing a set of real numbers that specify the coordinates of the particles. A nice feature of the QMEP model is that a set of integers fully gives the configuration, the collection $(\ell_{ij})_{(i,j) \in \Lambda}$. Here $\ell_{ij}$ specifies the local branch of cell $(i,j)$ and also counts the number of slips experienced by each cell $(i,j)$,  relative to the freshly prepared untrained glass for which $\ell_{ij} = 0$.
It is thus easy to define a stroboscopic distance between two configurations $T$ and $R$ as a Hamming distance between their elastic branch configurations  $T = (\ell^T_{ij})_{(i,j) \in \Lambda}$ and $R = (\ell^R_{ij})_{(i,j) \in \Lambda}$ as 
\begin{equation}
  d_{st}(R,T) = \frac{\left \vert \left \{ (i,j) \in \Lambda :  \ell^R_{ij} \ne \ell^T_{ij} \right \} \right \vert}{\vert \Lambda \vert},
  \label{stroboscopic_distance}
\end{equation}
where $\vert \Lambda \vert = N^2$ is the number of cells of the mesoscale model, and $\vert A \vert$ counts the number of elements of a set $A$. 
The distance $d_{st}(R,T)$ thus varies between $0$ for identical configurations
and $1$ for configurations so distinct that they do not overlap on any
of the cells.

\begin{figure}[t!]
\begin{center}
\includegraphics[width=0.50\columnwidth]{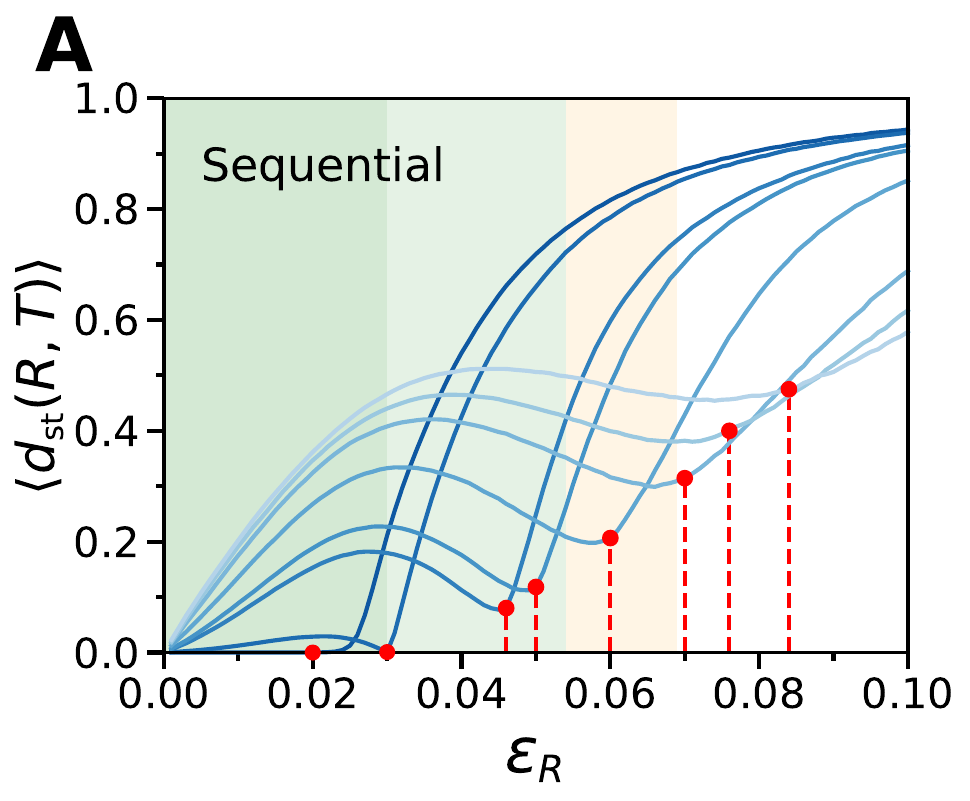}
\hspace{-10pt}
\includegraphics[width=0.50\columnwidth]{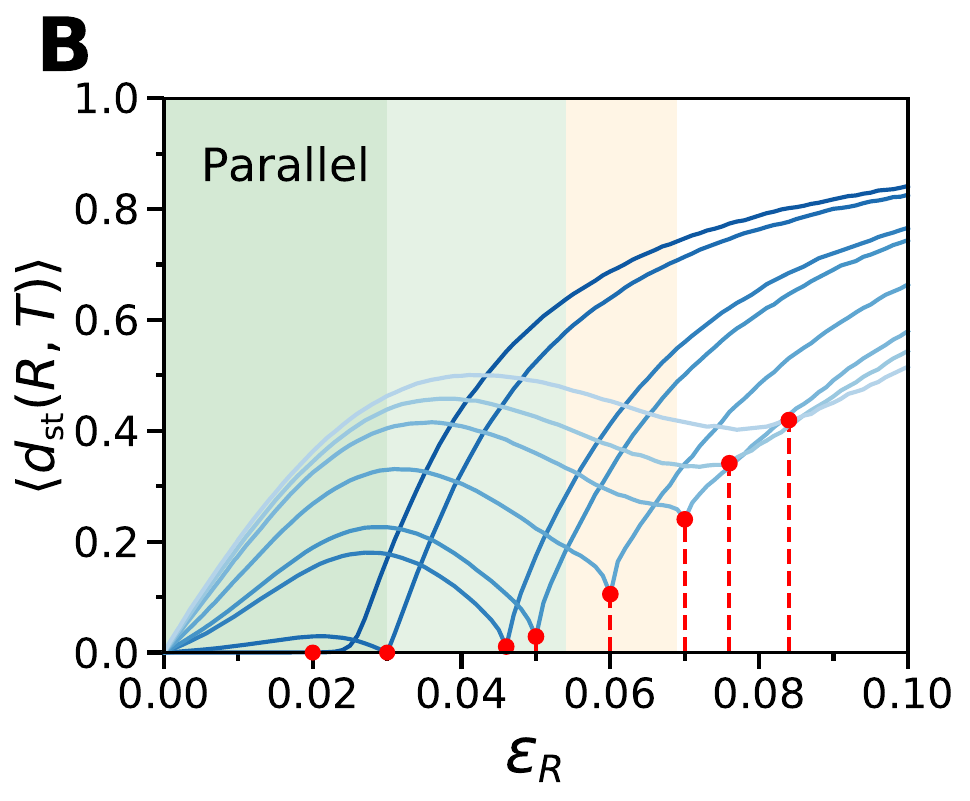}
\includegraphics[width=0.50\columnwidth]{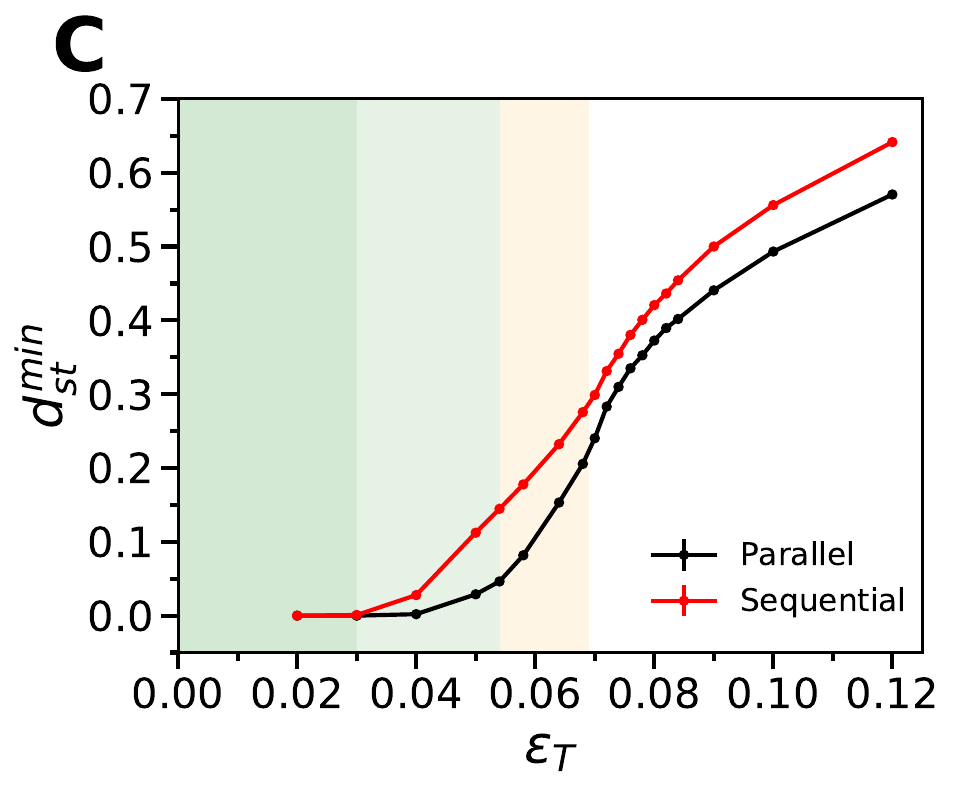}
\hspace{-10pt}
\includegraphics[width=0.50\columnwidth]{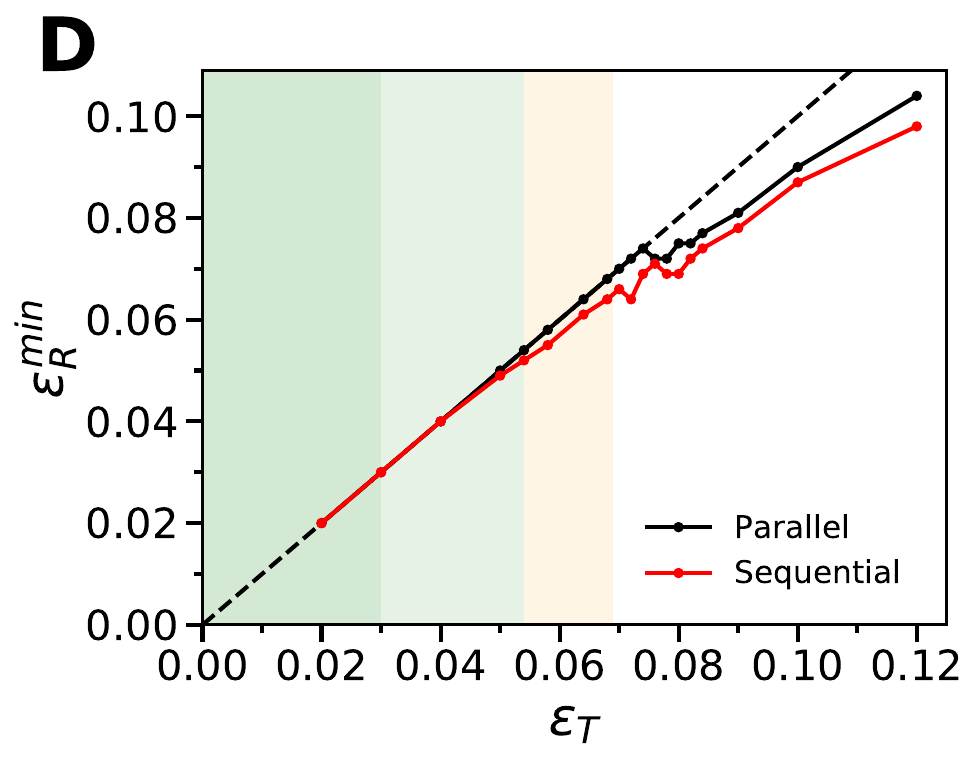}
\end{center}
 \caption{{\bf Memory response of the QMEP model after oscillatory shear training at strain amplitude $\varepsilon_{T}$.} Read-out response for A sequential and B parallel read-out protocol; C Evolution of the minimum stroboscopic distance $d_{strobo}^{min}$ over a read-out protocol with the training amplitude $\varepsilon_T$; D Evolution of the measured amplitude $\varepsilon_m$ at the minimum stroboscopic distance $d_{st}^{min}$ with the training amplitude $\varepsilon_T$. Background colors indicate the different regimes of reversibility. Dark green: all realizations lock into a monoperiodic cyle; light green: some of the limit cycles are multiperiodic; orange: some realization do not reach a limit cycle; grey: above irreversibility transition.} 
  \label{Memory-amplitude}
\end{figure}

In Fig.~\ref{Memory-amplitude}A and B, we show the evolution of the stroboscopic distance with read-out amplitude $\varepsilon_R$ for a
set of trained states and the cases of sequential and parallel read-out protocols. The training amplitudes $\varepsilon_T$ chosen vary from $0.02$ to $0.085$ and have been marked by vertical dashed lines whose termination on the corresponding read-out curve is indicated by a red circle. 
For the lowest training amplitudes, $\varepsilon_T = 0.02$,  the read-out response $d_{st}(R,T)$  vanishes for $\varepsilon_R \lesssim \varepsilon_T$, and the resulting response cycles is perfectly elastic, with all of the cells remaining in their local elastic branches. At higher read-out amplitudes, the curves exhibit a non-monotonic behavior of $d_{st}(R,T)$ with $\varepsilon_R$ developing a clear local minimum close to the value of the training amplitude. The same observations also hold in the parallel-read-out protocol case, shown in Fig.~\ref{Memory-amplitude}B.
Observe that since in the case of parallel read-out the shear cycles are applied to replicas of the trained state $T$,  there is no accumulation of stroboscopic distances due to the prior read-out cycles, as is the case for sequential read-outs. We believe this is why minima observed under parallel read-out are sharper and more clearly defined. Our QMEP model thus nicely reproduces the memory features recently observed in particle simulations and experiments~\cite{keim2011generic,Fiocco-PRL14, adhikari2018memory,mukherji2019strength,keim2020global,benson2021memory, arceri2021marginal,keim2022ringdown, shohat2023dissipation}. 

As illustrated in Fig.~\ref{Memory-amplitude}C, for both sequential and
parallel read-out, we observe that the value $d_{st}^{min}$ of the local minimum of the stroboscopic distance $d_{st}$ departs from zero and gradually increases with
the training amplitude. The background colors refer to different
regimes identified in Ref.~\cite{Kumar-PhD23} and give us a guide to
better understand the evolution of $d_{st}^{min}$. The leftmost region, shaded in dark green, indicates the range of training amplitudes where the system attains limit-cycles, and their response is monoperiodic. Hence, we expect that $d_{st}^{min}=0$ in this regime. The middle region, shaded in light green, indicates the range of training amplitudes where all systems attain limit-cycles, but where some of these are multiperiodic, i.e. their period spans two or more driving cycles. The slow increase of $d_{st}^{min}$ in this range is entirely due to the gradual
emergence of complex limit cycles of longer and longer
periods~\cite{Khirallah-Maloney-PRL21,Kumar-JCP22}. In the case of multiperiodic response, since the read-out protocol consists of a single cycle only, the trained and read configurations are no longer identical. Finally, the right region, shaded in orange, indicates the range of training amplitudes where some samples fail to attain cyclic response during the prescribed duration of $\mathcal{N}_T$ applied training cycles. Here, we find that the fraction of trained samples attaining cyclic response decreases with increasing $\varepsilon_T$. The upper limit of this regime is marked by the onset of the irreversibility transition $\varepsilon_{irr}$, which we had defined earlier as the training amplitude at which half of the
samples reach a limit cycle. Beyond this regime, a fraction of the samples still manages to attain cyclic response, so a shallow minimum is still observable outside of it.


Lastly, we quantify the precision with which the
sequential and parallel read-out protocols can infer the trained strain $\varepsilon_T$. Denote by $\varepsilon_{R}^{min}$  the value of the strain at the local minimum of the read-out $d_{st}(R,T)$ versus $\varepsilon_R$ curves, such as those shown in Fig.~\ref{Memory-amplitude}A and B. In Fig.~\ref{Memory-amplitude}D we plot $\varepsilon_{R}^{min}$ against the trained
amplitude $\varepsilon_T$. The dashed line $\varepsilon_{R}^{min} =
\varepsilon_T$ serves as a guide to the eye. We observe that 
for parallel read-outs (black symbols) and up to
the irreversibility transition, $\varepsilon_{R}^{min}$ furnishes a rather precise estimate of the training amplitude. This behavior extends to even slightly larger values of $\varepsilon_T$. In contrast, for the sequential read-outs, the precision of
the measurement gradually degrades as the training amplitude
increases. 

\subsection*{Fluctuations of the read-out response unveils the irreversibility transition}

\begin{figure}
\includegraphics[width=0.99\columnwidth]{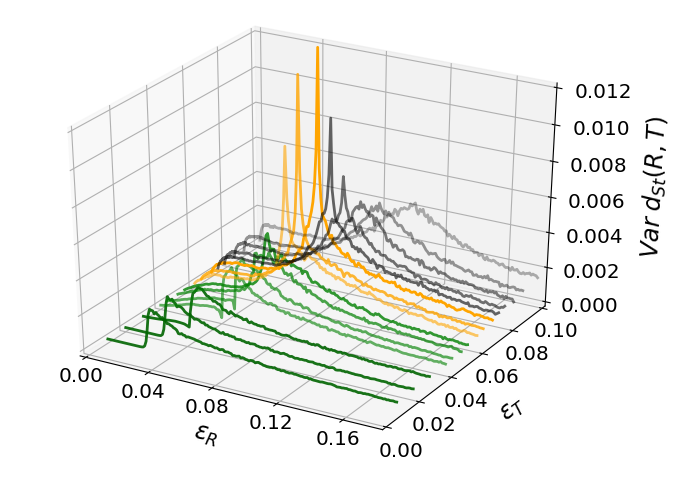}
  \caption{{\bf Sample-to-sample fluctuations at read-out.} Variance of the read-out response over an ensemble of
    1000 realizations in its dependence on read-out amplitude $\varepsilon_R$ for
    various training amplitudes $\varepsilon_T$. For the different training amplitudes shown, the variance exhibits a maximum when $\varepsilon_R \approx \varepsilon_T$. The global maximum of the variance is attained near the strain marking the onset of the irrversibility transition,  $\varepsilon_R \approx \varepsilon_T\approx 0.07$. Line colors reproduce the choice background colors described in Fig.~\ref{Memory-amplitude} and  label the different regimes of reversibility.} 
  \label{Max_fluctuations}
\end{figure}

We now discuss the sample-to-sample fluctuation behavior of the memory read-out. For simplicity, we restrict ourselves to the case of parallel read-outs. In Fig.~\ref{Max_fluctuations} we show the variance $V(\varepsilon_T,\varepsilon_R)$ of the stroboscopic distance $d_{st}(R,T)$ under parallel in-phase read-outs and for an ensemble of samples prepared at training amplitudes $\varepsilon_T$ with subsequent read-outs at $\varepsilon_R$. Note that since by construction $d_{st}(R,T) \in [0,1]$, the maximum variance (corresponding to a bimodal distribution with half of the samples at $d_{st}=0$ and the other half at $d_{st}=1$) is $V_{max}=0.25$.

We start with the behavior of $V(\varepsilon_T,\varepsilon_R)$ with $\varepsilon_R$ when $\varepsilon_T$ is small. From Fig.~\ref{Max_fluctuations} we see that for $\varepsilon_T=0.03$, the variance $V(\varepsilon_T,\varepsilon_R)$ exhibits a minimum when $\varepsilon_R=\varepsilon_T$ and actually vanishes there. As can be seen from  Fig.~\ref{Memory-amplitude}B, the vanishing of $V$ at  $\varepsilon_R=0.03$ simply reflects the fact that for low training amplitudes, all realizations are locked into monoperiodic limit cycles. Increasing $\varepsilon_R$ further, the variance first peaks just after $\varepsilon_T$ and then decreases gradually. This decrease is related to the bounded character of the stroboscopic distance $d_{st}$. For large values of $\varepsilon_R$, the stroboscopic distance gradually approaches its upper bound. Hence, sample-to-sample fluctuations get smaller again, and the variance $V(\varepsilon_T,\varepsilon_R)$ decreases.

Interestingly, the behavior of $V(\varepsilon_T,\varepsilon_R)$ with $\varepsilon_R$ gets significantly modified when the
training amplitude $\varepsilon_T$ is increased: the sharp minimum at
$\varepsilon_R \approx \varepsilon_T$ gradually disappears and becomes 
an increasingly sharp global maximum of $V(\varepsilon_T,\varepsilon_R)$ when $\varepsilon_R \approx \varepsilon_T$ is around $\varepsilon_{irr}$. When $\varepsilon_T$ 
is increased further, the maximum of $V$ (as a function of $\varepsilon_R$)  starts to broaden again. The change of behavior of  $V(\varepsilon_T,\varepsilon_R)$ along the line 
$\varepsilon_R=\varepsilon_T$,   reflects the increasing complexity and
the gradual disappearance of the limit cycles at high training
amplitude. Upon increasing the training amplitude, more and more
realizations reach a multi-periodic limit cycle or do not reach a
limit cycle at all so that the distribution of the distance
$d(\varepsilon_T,\varepsilon_T)$ gets bimodal and its variance
increases.

For a given training amplitude $\varepsilon_T$, denote by $V_{max}(\varepsilon_T)$ the maximum variance attained as the read-out amplitude is changed. 
As already noted above, we observe a well-defined peak of $V_{max}(\varepsilon_T)$ near the irreversibility transition, i.e. when  $\varepsilon_T \approx \varepsilon_{irr}=0.0689$) (see also Fig.~S1 in the SI). The
emergence of this maximum of fluctuations stems directly from the
definition of the irreversibility transition. At $\varepsilon_T
= \varepsilon_{irr}$ half of the realization lock-in into a limit
cycle while the other half does not. Therefore,  the stroboscopic distance
is almost evenly spread over a  bimodal
distribution of $d_{st}(R,T)$ whose low (high) values indicate  the presence (absence) of limit cycles. This result demonstrates
that the irreversibility transition can be directly identified from the
sample-to-sample fluctuation behavior of the read-out protocol.

\section*{Emergence of training induced structural anisotropy}




\begin{figure*}[t!]
\begin{center}
  \includegraphics[width=0.32\textwidth]{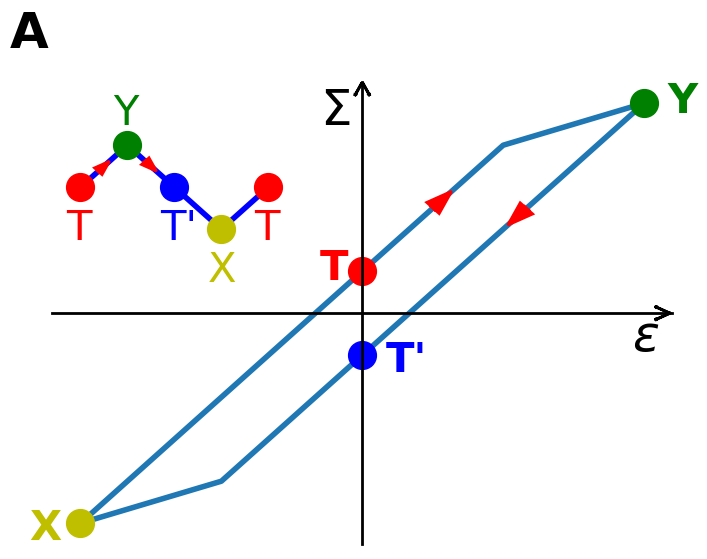}
  \includegraphics[width=0.315\textwidth]{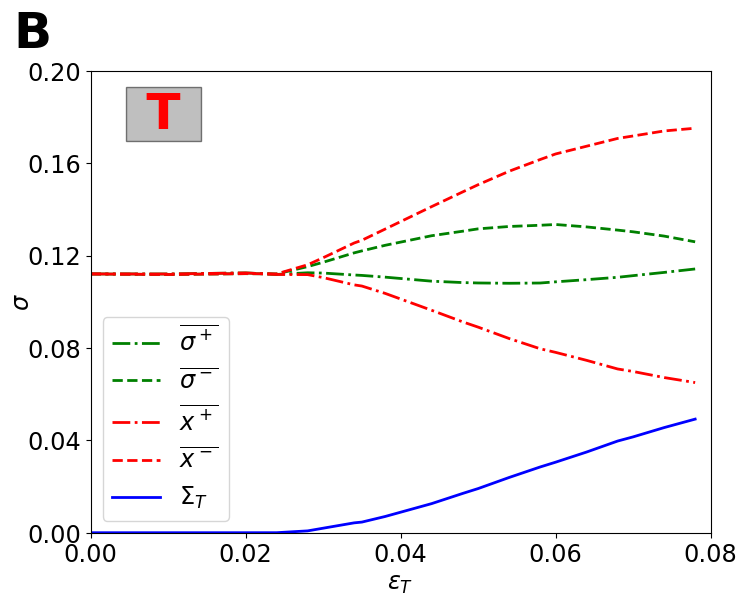}
  \includegraphics[width=0.33\textwidth]{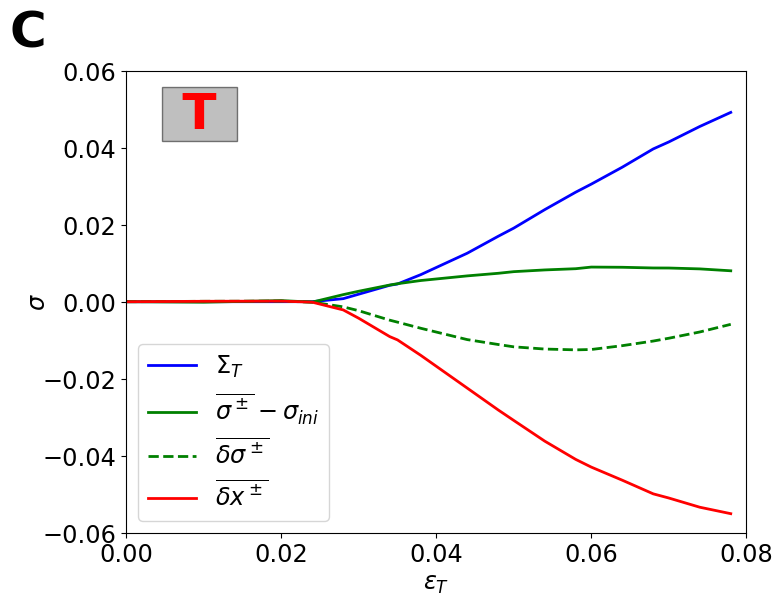}
  \includegraphics[width=0.32\textwidth]{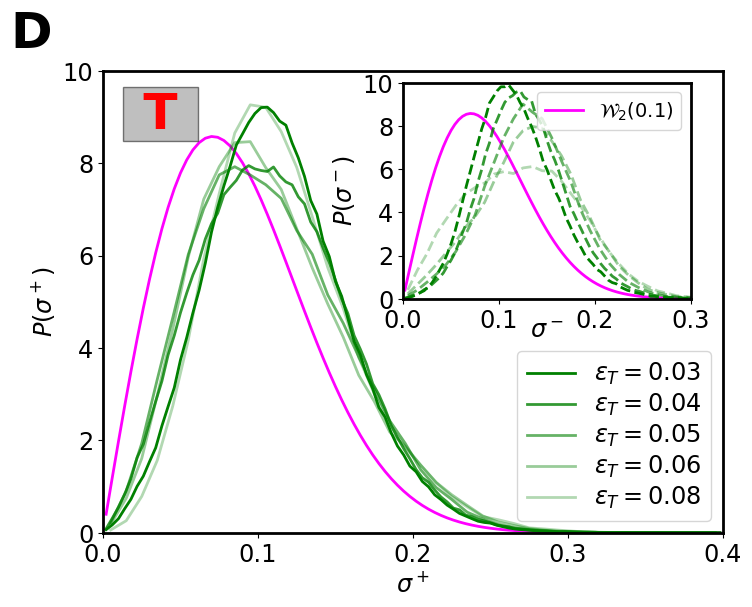}
  \includegraphics[width=0.32\textwidth]{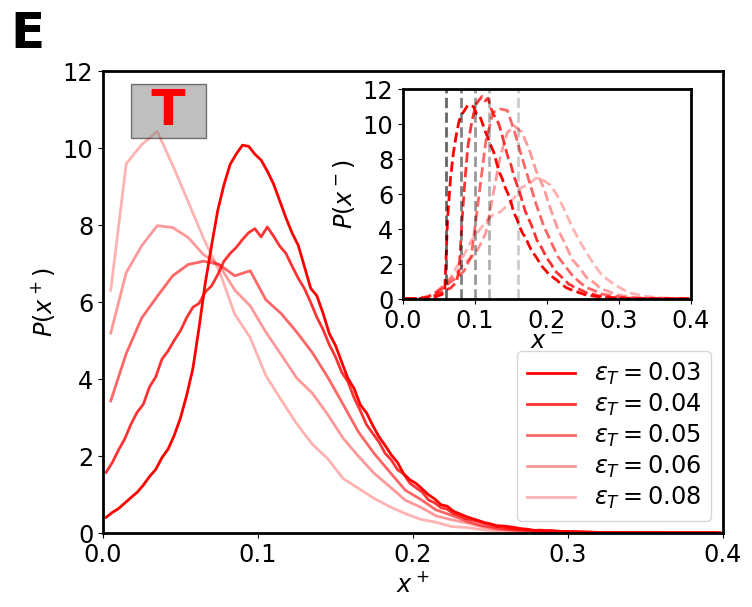}
  \includegraphics[width=0.32\textwidth]{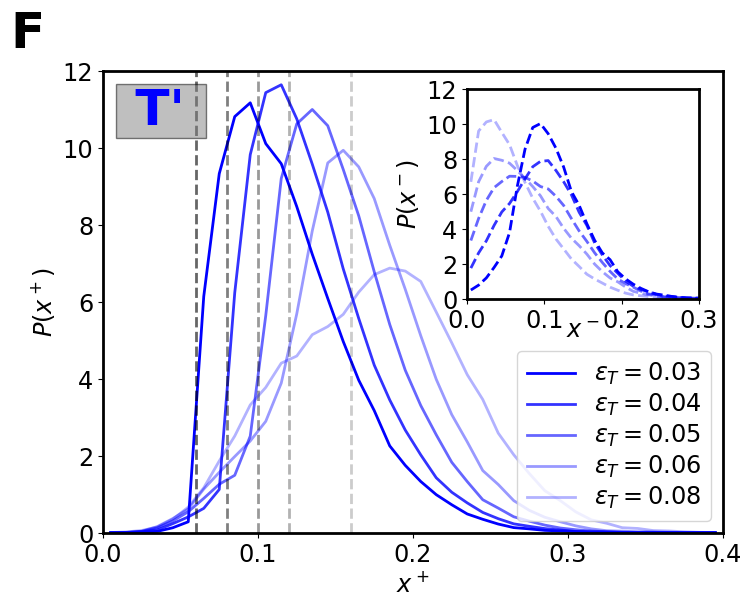}
\end{center}  
\caption{{\bf Emergence of anisotropy and parity as a result of training.} A: Sketch of the stress-strain relation of a limit-cycle under applied oscillatory shear $0 \to \varepsilon_T \to 0 \to  - \varepsilon_T \to 0$. The sense of orientation is indicated by the red arrows and the sequence of local elastic branch configurations visited during the cycle is $T \to Y \to T' \to X \to T$. 
B and C: The behavior with training amplitude $\varepsilon_T$ of sample-averaged properties of local elastic branches $\ell_{ij}$ of cells $(i,j)$ in the trained state $T$: B: Sample-averaged local stress thresholds $\overline{\sigma^\pm}$, plastic strengths $\overline{x^\pm}$ and residual stress $\Sigma_T$. C: Training-induced hardening of local elastic branches  relative to untrained sample, as quantified by $\overline{\sigma^\pm} = \overline{(\sigma_{ij}^+ + \sigma_{ij}^-)}/2$. The plot shows also the average anisotropies in local stress thresholds
$\overline{\delta \sigma^\pm} = \overline{(\sigma_{ij}^+ - \sigma_{ij}^-)}/2$ and plastic strengths 
$\overline{\delta x^\pm} = \overline{(x_{ij}^+ - x_{ij}^-})/2$. Note that all four quantities coincide up to the hysteresis transition $\varepsilon_{hys} = 0.025$, marking the onset of plasticity.
D: The main and inset show the distribution of forward local stress thresholds $\sigma^+$, respectively $\sigma^-$ at $T$. E and F: The distributions of local plastic strengths $x^+$ (main figure) and $x^-$ (inset) at states $T$ and $T'$ of the cyclic response. The dashed vertical lines are located at $2 \varepsilon_T$.}
\label{Polarization}
\end{figure*}

Shear plasticity induces stress and structure anisotropy. The existence of a residual strain classically defines plasticity as a result of loading a system and subsequently unloading it back into a stress-free condition. 
The return to an unstrained state then requires the application of a non-zero global stress. Plasticity within a training cycle (hysteresis) is thus expected to induce stress anisotropies.  
Interestingly, in addition to this {\it macroscopic} effect, shear plasticity also induces similar structural anisotropies at the {\it microscopic} scale. These have been observed in the alignment of polymer chains, the anisotropy of networks of granular contacts~\cite{Radjai-inbook04}, as well as in silica atoms~\cite{RVTBR-PRL09,sato2013differential,wakabayashi2015enhanced}, and can be quantified by nematic order parameters. 
At {\it mesoscopic scales}, the emergence of plasticity-induced anisotropies can be probed by the directional polarization of the distribution of local yield stresses~\cite{patinet2020origin,Patinet-CRPhys21}: local yields stresses tend to be higher in one direction than in the other one.

Fig.~\ref{Polarization}A is a sketch of a monoperiodic stress-strain limit cycle under oscillatory shear. As the applied shear strain varies as $0 \to \varepsilon_T \to 0 \to - \varepsilon_T \to 0$, the system traces out a hysteresis loop given by the branch configurations $T \to Y \to T' \to X \to T$, as shown. We will denote the residual stresses at $T$ and $T'$ as $\Sigma_T$ and $\Sigma_{T'}$, respectively.

In Fig.~\ref{Polarization}B we show how the global stress $\Sigma$ as well as the average local stress thresholds $\overline{\sigma^{\pm}}$ and plastic strengths $\overline{x^{\pm}}$ at $T$ vary with training amplitude $\varepsilon_T$.  Averages have been performed over all cells (i,j) in the state $T$ and across $300$ realizations of glasses. All quantities coincide up to an amplitude $\varepsilon_T = \varepsilon_{hys} = 0.025$, up to which the response is purely elastic and hence the global stresses at $T$ and $T'$ vanish, i.e. $\Sigma_T = \Sigma_{T'} = 0$. The common value of $\sigma_{ini} = 0.112$, which we observe for the stress thresholds and the plastic strengths, is the mean value of the stress threshold distribution of the freshly-aged but not yet trained glasses. 

Once $\varepsilon_T$ becomes larger than $\varepsilon_{hys}$, a residual stress $\Sigma_T$ starts to build up. In parallel, the plastic events triggered by the driving lead to a {\em dynamical selection} of local elastic branches, leading to an average stress threshold that starts to systematically deviate from $\sigma_{ini}$. 
As apparent from Fig.~\ref{Polarization}B, this deviation depends on the sense of the threshold relative to the driving direction. At $T$, where the applied strain is increasing in the {\em forward} direction, $\overline{\sigma^-}$ and $\overline{\sigma^+}$ increase, respectively decrease, relative to $\sigma_{ini}$ with increasing $\varepsilon_T$. A similar behavior is observed for the mean plastic strengths $\overline{x^\pm}$, which is more pronounced, since $\overline{x^\pm}$ also depends on the residual stress: $\overline{x^\pm} = \overline{\sigma^\pm} \mp \Sigma_T$. We find that the mean plastic strengths $\overline{x^+}$ in the forward direction are smaller than those in the reverse direction. This implies that at $T$ the change in applied strain necessary to trigger a plastic event in the forward direction is smaller than in the reverse direction of shearing.

The contrasting distributions of forward and backward thresholds after training indicate the development of a structural anisotropy. In Fig.~\ref{Polarization}C, we focus in more detail on this training-induced hardening and polarisation behavior. We show the evolution with training amplitude of the mean stress threshold relative to its initial value, $ \overline{\sigma^\pm}-\sigma_{ini}$ (where $ \overline{\sigma^\pm} = (\overline{\sigma^+}+\overline{\sigma^-})/2$) as well as those of the polarizations of thresholds and plastic strengths: $\overline{\delta \sigma^\pm} = (\overline{\sigma^+}-\overline{\sigma^-})/2$ and $\overline{\delta x^\pm} = (\overline{x^+}-\overline{x^-})/2$.  
As a reference scale, we plot the residual stress's evolution $\Sigma_T$. As expected, we observe a polarisation effect: both $\overline{\delta \sigma^\pm}$ and $\overline{\delta x^\pm}$ depart from zero above the hysteresis transition. In addition, we also observe a significant hardening effect: the mean thresholds $(\overline{\sigma^+}+\overline{\sigma^-})/2$ tend to get higher upon training. In other words, the length of the local elastic branches (their stability range) increases. 
Interestingly, the evolution of both thresholds hardening and polarization is non-monotonic. A peak is observed around $\varepsilon_T\approx 0.55$, the value above which not all systems find limit cycles anymore. 
In contrast, due to its dependence on residual stress, the plastic strength polarisation  $\overline{\delta x^\pm} = \overline{\delta \sigma^\pm} -\Sigma_T$ keeps increasing with the training amplitude  $\varepsilon_T$.

Fig.~\ref{Polarization}D shows the distribution of the  stress-thresholds $\sigma^+$ in state $T$, whose averages $\overline{\sigma^+}$ have been depicted in Panel (B), for $\varepsilon_T= 0.03, 0.04, 0.05, 0.06$ and $0.08$. Note that the last value is above the irreversibility transition $\varepsilon_{irr} = 0.0689$. The distribution in purple is the a priori Weibull distribution of thresholds for the untreated glasses. The slight decrease of $\overline{\sigma^+}$ with $\varepsilon_T$ in (B) is seen to be predominantly due to an asymmetric broadening of the distributions. The distributions of $\sigma^-$,  presented in the inset of Fig.~\ref{Polarization}D, exhibit a rather different behavior. With increasing $\varepsilon_T$, the shapes of the distributions largely remain the same while their peak is moving systematically towards larger values. The dynamical selection process that gives rise to the distribution of stress thresholds in state $T$ thus appears to operate on the reverse thresholds $\sigma^-$ mainly.

The main panel and inset of Fig.~\ref{Polarization}E shows the distribution of plastic strengths $x^+$, respectively $x^-$ at $T$. The vertical dashed lines in the insets correspond to the values $x_T = 2\varepsilon_T$, the elastic stress associated with a shear strain $\varepsilon_T$.
Notice the increase of asymmetry of the distributions of $x^\pm$ at $T$ as $\varepsilon_T$ increases.
While the distribution of $x^+$ starts to have an increasing number of cells with low plastic strengths, the corresponding population of such cells in the distribution of $x^-$ gets increasingly lower. In particular, a sharp front emerges at $x_T$, below which only a residual population of weak sites persists (note that these $x^-$ values below $x_T$ gradually build up a generic distribution when $x_T$ increases).
This is a first hint of the emergence of a mechanical memory at the mesoscopic scale. The value of the training amplitude gets imprinted as a clear feature in the distribution of plastic strengths. Note that at the higher training amplitude $\varepsilon_T=0.08$ which is above $\varepsilon_{irr}$, the front at $x_T$ has almost entirely disappeared.

Fig.~\ref{Polarization}F shows the distribution of plastic strengths at the antipode $T'$ of the limit cycle where the applied strain is again zero but decreasing. The plastic strengths in the direction of shearing are now given by $x^-$. In fact, for the training amplitudes shown, the distributions of $x^\pm$ at $T$ and $x^\mp$ at $T'$ are statistically identical. Given that the full cycle of applied shear returns the system back to its initial strain, these findings imply that the branches $X \to T \to Y$ and $Y \to T' \to X$ under increasing and decreasing shear, respectively, are also statistically indistinguishable under reversal of shear directions. Thus, we have a parity operation $\pi$ that maps the statistical properties of the two branches and, hence, the corresponding antipodal states $T$ and $T'$ into each other. 
Let us emphasize that this parity {\em emerges} due to the cyclic shearing. Moreover, this parity persists even at large shear amplitudes where a cyclic response is not attainable anymore.

\begin{figure*}[t!]
\begin{center}
\begin{minipage}{0.35\textwidth}
 \includegraphics[width=\textwidth]{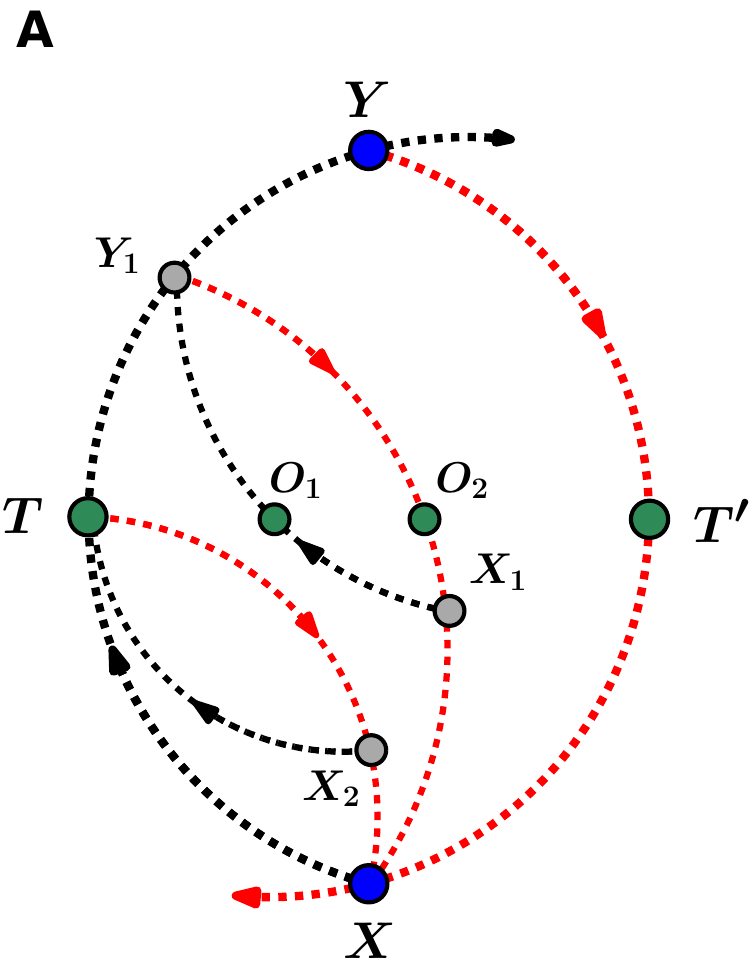}    
\end{minipage}
\hspace{-10pt}
\begin{minipage}{0.63\textwidth}
\includegraphics[width=0.49\textwidth]{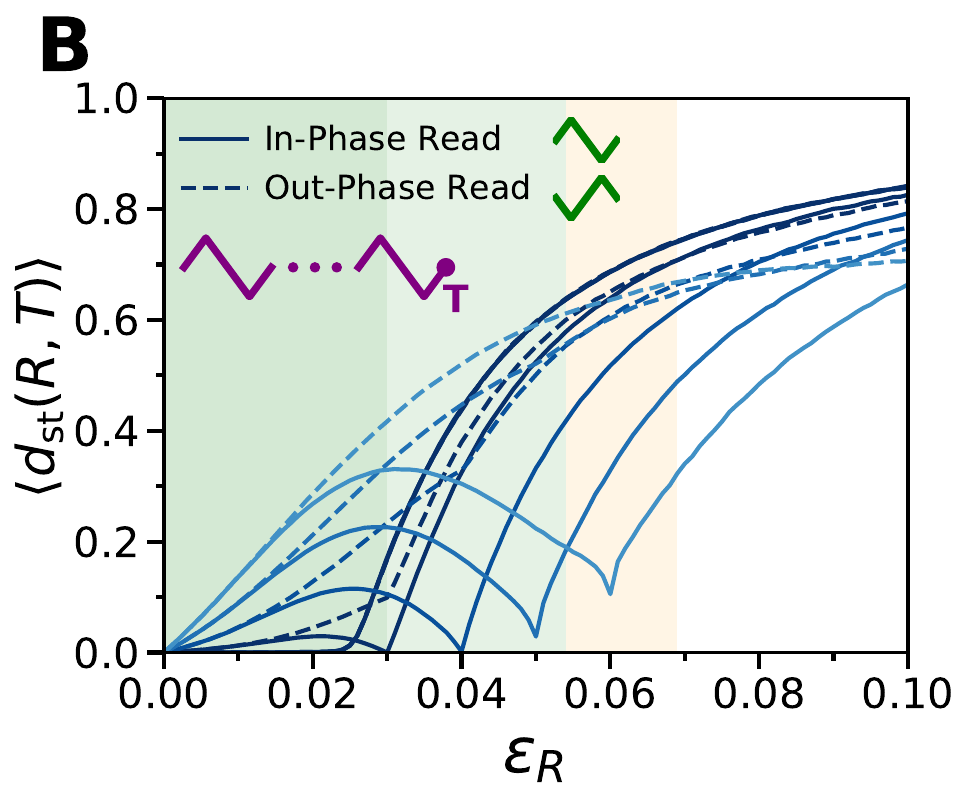}
\includegraphics[width=0.49\textwidth]{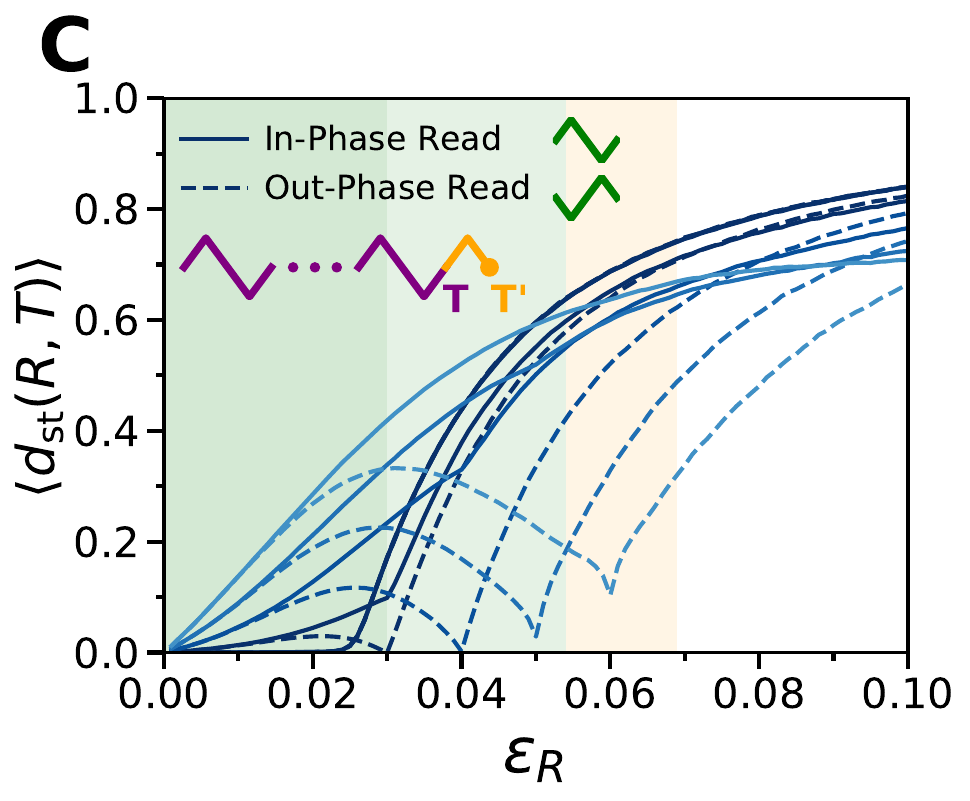}
\vspace{-10pt}
\includegraphics[width=0.49\textwidth]{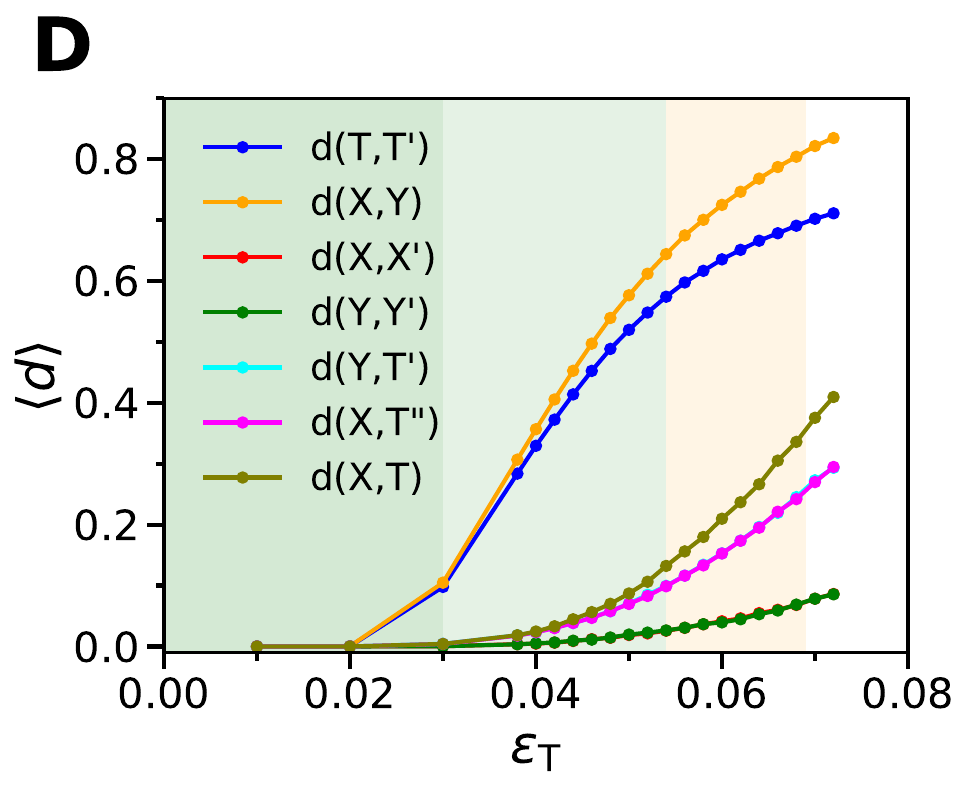}
\includegraphics[width=0.49\textwidth]{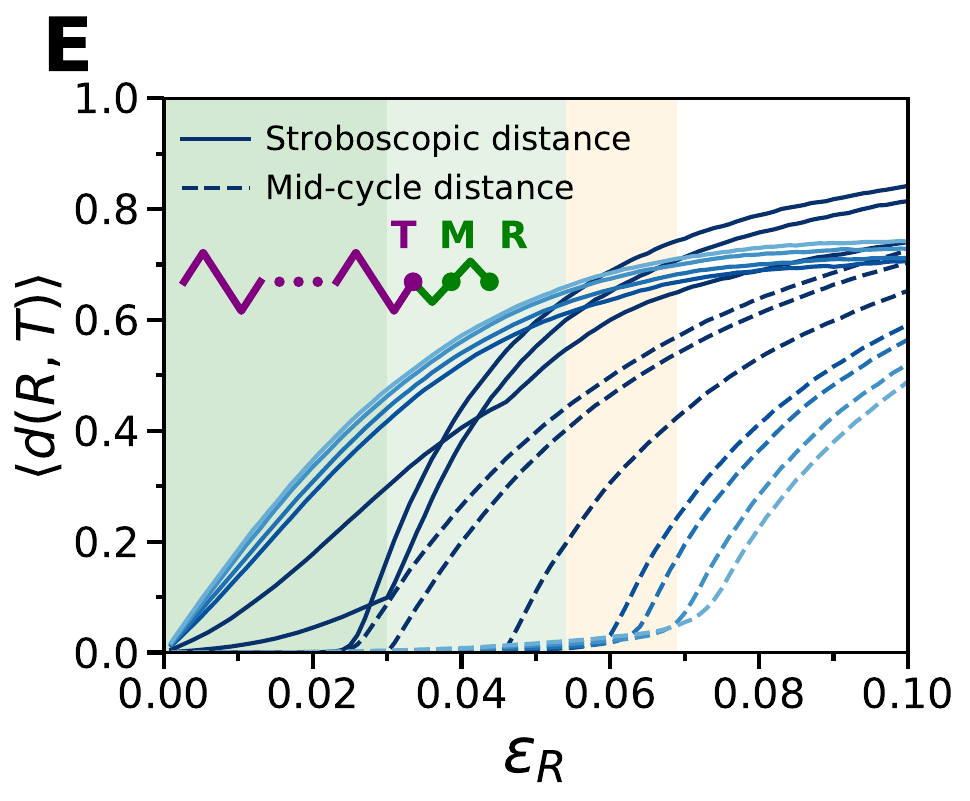}
\end{minipage}
\end{center}
\caption{{\bf Emergence of return-point memory and parity of shear direction.} A: Sketch of the transition between branch configurations of the QMEP model forming a limit-cycle. When a cyclic shear protocol $0 \to \varepsilon_T \to 0 \to -\varepsilon_T \to 0$ is applied to $T$, the system transits through the branch configurations $T \to Y \to T' \to X \to T$, as shown. The sequence of transitions between configurations under increase (decrease) in strain is indicated by the black (red) dashed arrows. A single-cycle of {\em in-phase} read-out $0 \to \varepsilon_R \to 0 \to -\varepsilon_R \to 0$ might follow the trajectory $T \to Y_1 \to O_2 \to X_1 \to O_1$. 
B: The effect of applying to the trained state $T$ the parallel read-out $0 \to \epsilon\varepsilon_R \to 0 \to -\epsilon \varepsilon_R \to 0$ to $T$, either in-phase ($\epsilon = 1$) or out-of phase ($\epsilon = -1$). Shown are the results for training amplitudes $\varepsilon_T = 0.02, 0.03, 0.04, 0.05 $ and $0.06$. The in-phase read-outs (solid lines) exhibit a strong memory of training amplitudes; the out-of-phase read-outs (dashed lines) do not show strong signs of memory. 
C: Parallel in- and out-of-phase read-outs applied to state $T'$, reached from $T$ by adding an in-phase half-cycle. Contrary to B, here, the in-phase read-outs (solid lines) do not exhibit a strong memory of training amplitudes, while the out-of-phase read-outs (dashed lines) exhibit cusps at the training amplitudes. Modulo phase, the read-outs from $T'$ turn out to be both qualitatively and quantitatively similar to those applied to $T$ and depicted in B, supporting the antipodal nature of $T$ and $T'$. 
D: The evolution of the   "widths" $d(T,T')$ and heights $d(X,Y)$ of the limit-cycles with $\varepsilon_T$. The states  $Y'$ and $X'$ are respectively reached by applying in and out-of-phase quarter-cycles $0 \to \pm \varepsilon_T$ to $T'$ and $T$. The distances $d(Y,Y')$ and   $d(X,X')$ 
measure the extent to which return-point memory holds, which in turn would imply that $X = X'$ and $Y = Y'$. E: By measuring the read-out states $M$ at the end of the first half-cycle, the stroboscopic distance $d_m(T,M)$ obtained from an out-of-phase read-out applied to $T$ can reveal the training amplitude $\varepsilon_T$ while the full-cycle out-of phase read-out cannot. Under RPM, $M = T$ for $\varepsilon_R \le \varepsilon_T$, which in turn would imply $d_m(T,M) = 0$. 
}
  \label{Memory-direction}
\end{figure*}

Oscillatory training thus induces two complementary effects on the glass structure. First we observe a mechanical annealing effect: upon increasing training amplitude, the mean stress thresholds $\overline{\sigma^\pm}$  increase, the glass hardens and gets more stable. The hardening shows a maximum around the irreversibility transition. This observation is consistent with the observation of a minimum of energy at the irreversibility transition reported in Refs.~\cite{Sastry-PRX19,yeh2020glass,bhaumik2021role,sri2021mesoland, Mungan-Sastry-PRL21, parley2022meanfield,Liu-Ferrero-JCP22}. Second, a symmetry breaking emerges between the strain-free configurations of the increasing and decreasing branches of the hysteresis cycle. The distribution of stress thresholds $\sigma^\pm_{ij}$ and plastic strengths $x^\pm_{ij}$  show mirroring anisotropic distributions. In configuration $T$, forward stress thresholds $\sigma^+_{ij}$ and plastic strengths $x^+_{ij}$ are lower than their reverse counterparts: on the increasing branch, the glass is softer in the forward direction than in the reverse direction. The opposite statements apply in the strain-free configuration $T'$ of the decreasing branch of the hysteresis cycle. The development of this symmetry breaking is accompanied by the formation of a gap in the distribution of the plastic strengths in the direction reverse to that of the applied shear, while there is no such gap in the direction of shear. The width of the gap in the distribution of reverse plastic strengths is precisely $2\varepsilon_T$, the elastic stress associated with the training amplitude $\varepsilon_T$.
Such gaps arise by construction in mean-field models that assume that the cyclic response of the mesoscale blocks is purely elastic, implying that $x^\pm > x_T$ for each block \cite{sri2021mesoland, Mungan-Sastry-PRL21,parley2022meanfield}. Our results show that the interplay between forward and reverse plastic strength is more complicated, and that under cyclic response such gaps open and close as the driving changes direction.  
These features can be regarded as a microscopic imprint of the oscillatory training on the glass structure, encoding thereby mechanical memory.

\section*{Memory of the last direction of shear and emergence of return-point-memory-like behavior}

Fig.~\ref{Memory-direction}A represents the transitions between branch configurations of the QMEP model, once a cyclic response has been attained that is monoperiodic.
The sequence of transitions between configurations under increase (decrease) in strain is indicated by the black (red) dashed arrows. As the cyclic shear protocol  $0 \to \varepsilon_T \to 0 \to -\varepsilon_T \to 0$ is applied to $T$, the system transits through the branch configurations $T \to Y \to T' \to X \to T$. A single-cycle of read-out $0 \to \varepsilon_R \to 0 \to -\varepsilon_R \to 0$ applied to $T$ with amplitude $\varepsilon_R < \varepsilon_T$, might follow the trajectory $T \to Y_1 \to O_2 \to X_1 \to O_1$, shown in Fig.~\ref{Memory-direction}A. At $Y_1$ and $X_1$ the directions of shear are reversed, and such states are denoted as {\em switch-back} states. For the moment, let us note that if return-point memory (RPM) were present,  continuing from $X_1$ by starting to increase the strain beyond $O_1$, the system would eventually reach the prior switch-back state $Y_1$, before continuing on the pathway leading to Y and beyond~\cite{munganterzi2018, mungan2019networks}. Likewise, with $T$ and $X_2$ representing a pair of consecutive switch-back states,  if RPM were to hold, increasing the strain at $X_2$ would return the system first to the previous switch-back state $T$. 

Defining the {\em in-phase} read-out protocol as $0 \to \epsilon \varepsilon_R \to 0 \to -\epsilon \varepsilon_R \to 0$, where $\epsilon = 1$, we  will refer to the case when $\epsilon = -1$  as an {\em out-of phase} read-out.  
Fig.~\ref{Memory-direction}B shows the result of applying parallel in- and out-of-phase read-outs to states $T$ of samples trained at amplitudes $\varepsilon_T = 0.02, 0.03, 0.04, 0.05$ and $0.06$. As we have already seen in Fig.~\ref{Memory-amplitude}B, the in-phase read-outs (solid lines) exhibit a strong memory of the training amplitude, which is evidenced by the pronounced local minima of $d_{st}(R,T)$ where the read-out amplitude matches the training amplitude. The out-of-phase read-outs (dashed lines) exhibit a memory that is revealed by a change of slope of $d_{st}(R,T)$ around $\varepsilon_R \approx \varepsilon_T$ but gets weaker with increasing training amplitude.

Fig.~\ref{Memory-direction}C: shows the result of applying parallel in- and out-of-phase read-outs to the state $T'$, which is reached from $T$ under the application of an additional (in-phase) half-cycle $0 \to \varepsilon_T \to 0$, cf. Fig.~\ref{Memory-direction}A. We see that its now the in-phase read-outs which do not bear a strong memory of the training, while the out-of-phase read-outs clearly do. Moreover, modulo read-out-phase, these two sets of read-out curves are rather similar both qualitatively and quantitatively. This finding is consistent with the notion of parity and the states $T$ and $T'$ being antipodal, as discussed in the previous section. Thus, it appears that the notion of parity applies not only to the distributions of local stress thresholds and plastic strengths, which are bulk quantities but also to families of microscopic deformation pathways during read-outs. Qualitatively similar behavior is observed for sequential read-out, and the results are shown in the SI.

Note that the read-outs at $T$ or $T'$ will recover the memory of the training amplitude only if we apply the correct read-out phase. On the other hand, the statistical invariance under the exchange of $T$ with $T'$ in combination with a switch of shearing direction implies that the memory can only be retrieved by a read-out if it is in phase with the last direction of shear. In fact, we can consider read-outs from a state $\widetilde{T}$ which we obtain by subjecting the trained state $T$ to an (additional) out-of-phase half-cycle $0 \to -\varepsilon_T \to 0$, so that the system transits through the states $0 \to X' \to \widetilde{T}$. 
This additional half-cycle does not change the last direction of shearing, and hence, we would expect an in- and out-of-phase read-out similar to the one from $T$. The SI contains the results of in-phase and out-of-phase read-outs from $\widetilde{T}$ and confirms this expectation. Upon closer inspection, we find that the branch configuration $X'$, which is reached at strain
$-\varepsilon_T$,  is very close to $X$. A subsequent increase of shear strain to zero leads to a state $\widetilde{T}$ close to $T$. Fig.~\ref{Memory-direction}D is a plot of $d(X,X')$ against $\varepsilon_T$, showing that it remains small over a large range of training amplitudes.

We can harness these observations to construct a read-out protocol that will infer the training amplitude without knowing the last direction of shear, or equivalently, irrespective of whether we were given for read-out the state $T$ or its antipode $T'$. To do so, we arbitrarily choose a direction of shear and then perform our read-out as before, but this time, we also record the mid-cycle state $M$ reached after applying the first half-cycle. Denoting again by $R$ the state reached at the end of the full read-out cycle, we record both the distances $d_m = d(T,M)$ and $d_{st} = d(T,R)$. 
Fig.~\ref{Memory-direction}E shows the read-out curves for training amplitudes $\varepsilon_T = 
0.02, 0.03, 0.046, 0.06, 0.064, 0.070$ and $0.074$, with 
the read-out chosen to be {\em out-of phase} relative to the sense of driving. As expected, the stroboscopic distance $d_{st}(T,R)$ does not permit reliable memory retrieval. However, the mid-cycle distance $d_m = d(T,M)$ remains very close to zero for read-out amplitudes $\varepsilon_R \le \varepsilon_T$ and then starts to rise abruptly. Moreover, this behavior persists for larger training amplitudes up to the irreversibility transition. 
Referring again to Fig.~\ref{Memory-direction}A we see that this behavior is consistent with RPM. As we have seen before, the out-of-phase half-cycle at amplitude $\varepsilon_R \le \varepsilon_T$ will give rise to a trajectory $T \to X_2 \to T$, since $T$ is a switch-back state under $X \to T \to X_2$. Thus, under RPM, we expect that $d_m = 0$ for $\varepsilon_R \le \varepsilon_T$. For $\varepsilon_R > \varepsilon_T$, the first-half cycle will cause the trajectory to go beyond $X$ during its initial segment of decreasing strain (red dashed line) and will thereby wipe out the switch-back state $X$ that was established as part of the training. Finally, let us note that the protocol with mid-cycle read-out can also be implemented sequentially and leads to similar results, as shown in the SI. Such sequential protocols can easily be implemented experimentally and thus will give access to both the amplitude and the last direction of shear during training.

In order to probe the RPM-like behavior further, we consider the states $X'$ and $Y'$ reached when applying to the antipodes $T$ and $T'$ the quarter phase cycles  $0 \to -\varepsilon_T$, and $0 \to \varepsilon_T$, respectively. Referring to Fig.~\ref{Memory-direction}A, RPM would imply $X = X'$ and $Y = Y'$, so that $d(X,X') = d(Y,Y') = 0$. Fig.~\ref{Memory-direction}D shows 
the behavior of $d(X,X')$ and $d(Y,Y')$  over a range of amplitudes $\varepsilon_T$. Observe, that 
$d(X,X') \approx d(Y,Y')$ over the full range of training amplitudes shown, extending even beyond the irreversibility transition. Moreover, the values of $d(X,X')$ and $d(Y,Y')$ remain close to zero for strain amplitudes well inside the regime of multiperiodic response (light green region), beyond which they start to increase rather slowly.

We have already noted that the distributions of plastic strengths of $x^-[T]$ and its antipode $x^+[T']$ are statistically indistinguishable and exhibit a gap for values $x^\pm < x_T$, i.e. $P(x^-[T] < x_T) \approx P(x^+[T'] < x_T) \approx 0$. This suggests that a strain increase (decrease) by $\varepsilon_T$ at $X$ ($Y$) will yield a predominantly elastic response.  Denote the sequence of states visited when a single in-phase shear strain cycle of amplitude $\varepsilon_T$ is applied to $T$ as $T \to Y \to T' \to X \to T''$, where $T = T''$ if the response is monoperiodic. The purple and cyan curves in Fig.~\ref{Memory-direction}D show how $d(X,T'')$ and  
$d(Y,T')$ change with $\varepsilon_T$.  We see again that their behavior is statistically indistinguishable, as expected by the parity of antipodal states. For the monoperiodic regime of training amplitudes (darker green region), these values remain close to zero and start to increase in the multiperiodic regime, implying that the transitions $X \to T''$ and $Y \to T'$ on the limit-cycle are nearly elastic for sufficiently low training amplitudes. 

To summarize, we have shown that a protocol that performs read-outs  at mid- and full-cycle and compares these with the  trained states, successfully retrieves directional memory, namely the last direction of the applied oscillatory shear as well as its amplitude. Its capability to do so relies on the RPM-like behavior that is exhibited by the trained glasses.

\section*{A Preisach-like model of directional memory}

We now use the results that we established in the two previous sections to build a minimal model of directional memory.
Assuming that the transitions from $X \to T''$ and $Y \to T'$ are perfectly elastic allows us to formulate a prediction of the in- and out-of-phase stroboscopic read-out distances, $d^{in}_{st}$ and $d^{out}_{st}$, cf. \eqref{stroboscopic_distance},  which we will base entirely on the distribution of $x^+[T]$, the forward plastic strengths at $T$. 

We consider first the case of an out-of-phase read-out, 
$0 \to -\varepsilon_R \to 0 \to \varepsilon_R \to 0$ and let $x_R = 2  \varepsilon_R$. By assumption, the first half of the read-out cycle is perfectly elastic and hence returns the system to $T$. In the second half of the cycle, all sites 
$(i,j)$ for which $x^+_{ij}[T] \le x_R$, are candidates to yield during the strain increase $0 \to \varepsilon_R$. Due to avalanches, some additional sites may be destabilized, while owing to the quadrupolar nature of stress redistribution, some of the initial candidate sites may be stabilized by the yielded sites. We will ignore these two types of sites and assume that the set of sites $(i,j)$ that yielded during the strain increase is precisely those for which $x^+_{ij}[T] \le x_R$. In addition, we will assume that the response to the final strain reduction $\varepsilon_R \to 0$ is purely elastic. This leads to the prediction that 
\begin{equation}
  d_{st}^{out}(R,T) = {\rm Prob}(x^+_{ij}[T] \le x_R).  
  \label{eqn:dst-out}
\end{equation}

\begin{figure}[t!]
\centering
\includegraphics[width=\columnwidth]{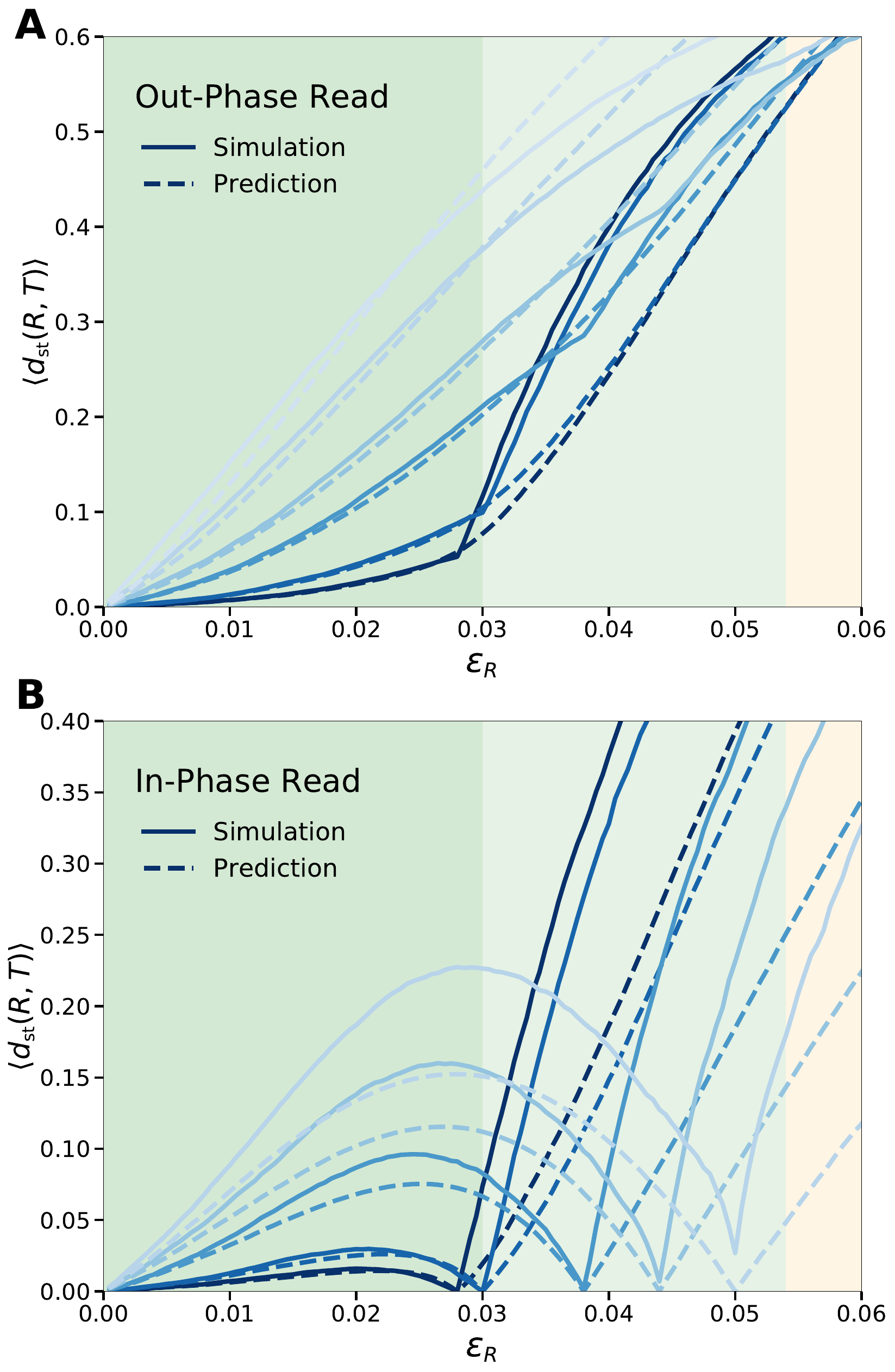}
\caption{ {\bf Simulation and theoretical response to the in- and out-of phase read-out protocol.} Comparison of the out-of-phase read-out response obtained from simulations (solid lines) with the theoretical prediction (dashed lines) based on the density of local sites susceptible to yielding, \eqref{eqn:dst-out}. The training amplitudes considered are $\varepsilon_T = 0.028, 0.03, 0.038, 0.044, 0.054$ and $0.064$ (from bottom to top). B: The in-phase read-out from simulations and compared with theoretical prediction for training amplitudes 
$\varepsilon_T = 0.028, 0.03, 0.038, 0.044$ and $0.05$ (bottom to top).  
}
\label{fig:in-out-sim-vs-theory}
\end{figure}

Fig.~\ref{fig:in-out-sim-vs-theory}A shows a comparison of the prediction of \eqref{eqn:dst-out},  using the numerically obtained distributions of $x^+_{ij}[T]$ shown in Fig.~\ref{Polarization}E, with the read-out results obtained from numerical simulations for training amplitudes $\varepsilon_T = 0.028, 0.03, 0.038, 0.044, 0.054$ and $0.064$.  The predictions of the lowest three training amplitudes agree rather well over a range $\varepsilon_R \lesssim \varepsilon_T$. For $\varepsilon_R > \varepsilon_T$, these predictions deviate significantly for the lowest two training amplitudes. At higher training amplitudes, a good agreement with the prediction is reached only for low values of read-out. 

A similar approach can be applied to obtain a prediction for the in-phase read-outs and leads for $\varepsilon_R \le \varepsilon_T$ to
\begin{align}
d_{st}^{in}(R,T) = &{\rm Prob}(x^+_{ij}[T] \le x_R) \nonumber \\ 
&\times {\rm Prob}(x_R < x^+_{ij}[T] \le x_T \,  \vert  \, x^+_{ij}[T] \le x_T),
\label{eqn:dst-in1}
\end{align}
while in the case that $\varepsilon_R > \varepsilon_T$, we have
\begin{equation}
 d_{st}^{in}(R,T) = {\rm Prob}(x_T < x^+_{ij}[T] \le x_R).
 \label{eqn:dst-in2}
\end{equation}

This result is based on a set of assumptions that we state next. Denoting by $F_T$ and $F_R$ the set of sites $(i,j)$ such that 
$x^+_{ij}[T] \le x_T$ and $x^+_{ij}[T] \le x_R$, respectively, we consider a sequence of configurations visited upon a read-out cycle whose transition pathway is $T \to Y_1 \to O_2 \to X_1 \to O_1$,  as sketched in Fig.~\ref{Memory-direction}A. We then assume that
\begin{itemize}
 \item[i.] the response after training is cyclic and monoperiodic,
 \item[ii.] the responses in the segments $Y_1 \to O_2$ and $X_1 \to O_1$ are purely elastic,
 \item[iii.] the set of sites that yield during the pathways $T \to Y_1$ is given by $F_R$.
\end{itemize}

Denoting by $F^-_R$ the set of sites that yield during the pathway $O_2\to X_1$, i.e.
\begin{equation}
F^-_R = \{ (i,j): x^-_{ij}[O_2] \le x_R \},
\label{eqn:FminusRdef}
\end{equation}
the above assumptions imply that $\vert F^-_R \vert =  \vert F_R \vert$ and  
\begin{equation}
  d_{st}^{in} = \frac{\vert F_R \triangle F^-_R \vert }{\vert \Lambda \vert}, 
  \label{eqn:dst_in_sets}
\end{equation}
where $A \triangle B$ is the symmetric set difference corresponding to the set of elements that are in $A$ but not in $B$ and vice versa.

In order to estimate $d_{st}^{in}$, we make the following two additional assumptions
\begin{itemize}
 \item[iv.]  
\begin{align}
  F^-_R &\subset F_T, \quad \varepsilon_R \le \varepsilon_T,  \nonumber \\ 
  F_T &\subset F^-_R, \quad \varepsilon_R > \varepsilon_T. 
\end{align}
 \item[v.] in the case that $\varepsilon_R \le \varepsilon_T$, the set $F^-_R$ is obtained by assigning to each site $(i,j) \in F_T$ a plastic strength value $x^-_{ij}[O_2]$ that is drawn independently and identically (iid) from the distribution of $x^+_{ij}[T]$ conditioned on $x^+_{ij}[T] \le x_T$. For $\varepsilon_R > \varepsilon_T$, $\vert F_T \vert$ sites are drawn iid as in the previous case, while the remaining sites are drawn iid, but conditioned on $x_T < x^+_{ij}[T] \le x_R$. 
\end{itemize}
Using assumption v. to replace the RHS of \eqref{eqn:dst_in_sets} by its average value then leads to \eqref{eqn:dst-in1} and \eqref{eqn:dst-in2}. Let us note that when $\varepsilon_R = \varepsilon_T$, our assumptions imply that $F_T = F_R = F^-_R$, so that from \eqref{eqn:dst_in_sets} it follows that $d_{st}^{in} = 0$.

Fig.~\ref{fig:in-out-sim-vs-theory}B compares the in-phase read-out distances $d_{st}^{in}$ from our simulations with the prediction obtained from \eqref{eqn:dst-in1} and \eqref{eqn:dst-in2} Shown are the results for training amplitudes $\varepsilon_T = 0.028, 0.03, 0.038, 0.044$ and $0.05$ (from bottom to top and in increasingly lighter shades of blue). Considering first the read-out regimes where $\varepsilon_R \le \varepsilon_T$, we see that 
the agreement with the predicted result is rather good for training amplitudes $\varepsilon_T$ in the region where the cyclic response is monoperiodic (region shaded in darker green). For all training amplitudes shown, the predicted asymmetric shape of $d_{st}^{in}$ matches well the numerical results. However, with increasing training amplitudes, the theoretical results systematically underestimate the read-out distances. Finally, in the region $\varepsilon_R > \varepsilon_T$, the sharp rise in $d_{st}^{in}$ is captured well by our predictions but the discrepancies increase with $\varepsilon_T$.

We conclude with some remarks. The choice of distribution described in assumption v. is motivated by our empirical observation of polarity, which implies that the distributions of plastic strengths $x^+[T]$ and $x^-[T']$ are identical. Hence, all that assumption v. does is to replace the distribution of $x^-[O_2]$ with that of $x^-[T']$.
The assumption iv. and v. effectively impose RPM by turning the theoretical description into a Preisach model \cite{preisach1935magnetische, terzi2020state}. Each site $(i,j) \in F_T$ emulates a Preisach hysteron, i.e. a unit of hysteresis,  which can hysteretically switch between two states. In the QMEP model, these hysterons emerge under training and correspond to the local elastic branches $\ell_{ij}$ at $T$ and $T'$, along with their switching fields which are determined by $x^+_{ij}[T]$ and 
$x^-_{ij}[T']$. Such a description ignores interactions between sites that arise when a given site yields, causing a redistribution of stresses and hence updates of plastic strengths $x^\pm$ at other sites. With larger training and read-out amplitudes, such interactions will be increasingly dominant, invalidating the assumptions underlying our description.

\section*{Conclusion}

To summarize, the study of our quenched mesoscopic elasto-plastic (QMEP) model of amorphous solids has allowed us to reproduce and understand the memory behavior observed both experimentally in dense suspensions, as well as numerically in particle simulations. We developed experimentally testable read-out protocols from which, besides the training amplitude, the direction of the last shear can be recovered.   The evolution of the read-out response upon increasing training amplitude shows that the memory gradually degrades due to the appearance of multiperiodic limit cycles and the gradual loss of plastic reversibility. Still, a memory behavior  persists even past the irreversibility transition. Interestingly, we find that the behavior of sample-to-sample fluctuations of the read-out response with training amplitude provides a direct means to identify the irreversibility transition.

The QMEP model, while being a minimal mesoscale model of an amorphous solid, thus allowed us to study in detail the mechanical mechanisms giving rise to memory formation under cyclic shear. In particular, we identified an emergent polarization behavior upon oscillatory training. In addition, our mesoscale model provides  direct access to structural mechanical information, such as the fields of local stresses, stress thresholds and plastic strengths, which allowed us to link features of memory formation to the mechanical annealing. These finding naturally  
motivate more detailed studies on the spatial support underlying the observed mechanical memory. A key feature of the QMEP model is the quenched character of the disorder: each local cell lives on a well defined frozen random landscape so that it may revisit the very same series of plastic thresholds upon forward and reverse shear. This strong assumption naturally motivates a more detailed characterization of the local mechanical disorder in atomistic simulations of glassy materials~\cite{PVF-PRL16,Richard-PRM20}.

Our results of encoding and subsequent read-out of the memory and direction of training amplitude reveal a behavior that is qualitatively consistent with what one would expect from systems that obey RPM. Building on this observation, we developed a simple Preisach-like model of directional memory, which qualitatively reproduces both the in-phase and the out-of-phase read-out responses. Note again that this quasi-RPM behavior, observed in experiments as well as numerical simulations \cite{keim2020global, mungan2019networks}, is somewhat unexpected and certainly deserves more scrutiny. 

Our findings in fact suggest that the   RPM-like behavior emerges as a result of mechanical annealing. \eqref{eqn:dst-out} shows that we can interpret $d_{st}^{out}$ as a proxy for 
a density of states, more specifically, the fraction of sites that can yield under a given read-out amplitude. Seen in this way, the change of slope in the empirical read-out distances $d_{st}^{out}$ around 
$\varepsilon_R \approx \varepsilon_T$ is due to the read-out causing the excitation of hitherto untrained degrees of freedom. The behavior of the mid-cycle distances with training amplitudes, shown in Fig~\ref{Memory-direction}E, further supports this interpretation. We plan to pursue these ideas in future work. 

The mesoscale model of an amorphous solid to cyclic shear
has provided a detailed description of the mechanical
processes underlying memory formation. We find that the driving leads
to the self-organization of local plasticity, which, via the evolution of stress thresholds under mechanical annealing, leads to the  
encoding of features of the driving history.

{Acnowledgements: The research of MM was funded by the Deutsche
  Forschungsgemeinschaft (DFG, German Research Foundation) under
  Projektnummer 398962893. MM would like to thank PMMH and ESPCI for
  their kind hospitality during multiple stays which were made
  possible by Chaire Joliot grants awarded to him. This project has
  received funding from the European Union’s Horizon 2020 research and
  innovation programme under the Marie Skłodowska-Curie grant
  agreement No. 754387. DV acknowledges insightful discussions with
  Craig Maloney and Elisabeth Agoritsas.}

\bibliography{Memory}

\newpage


\newpage
\appendix

\setcounter{figure}{0}
\renewcommand{\thefigure}{S\arabic{figure}}

\section*{Supplementary Information}

This SI contains the figures that describe additional results referred to in the main text. These were obtained from $16\times 16$ poorly-aged realizations of our QMEP model. 




\subsection*{Maximum variance of stroboscopic distances during read-out}

\begin{figure}[h!]
\includegraphics[width=\columnwidth]{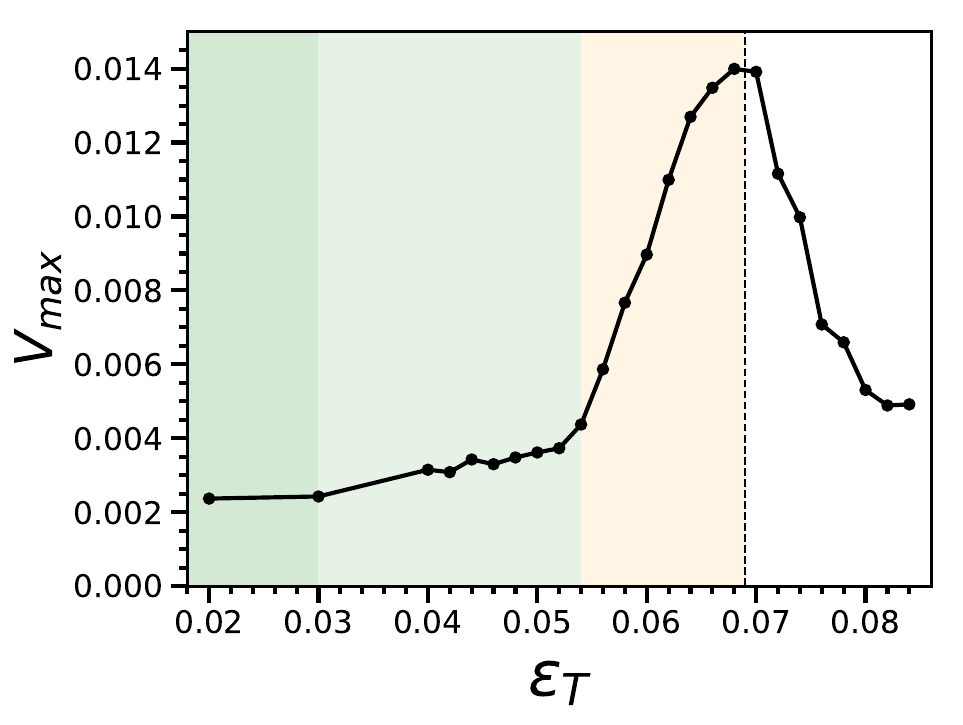}
\caption{Maximum of variance $V_{\rm max}(\varepsilon_T)$ of the stroboscopic distance $d_{St}(\varepsilon_R,\varepsilon_T)$ as a function of the training amplitude $\varepsilon_T$. $V_{\rm max}(\varepsilon_T)$ shows a maximum at  $\varepsilon_T \approx   \varepsilon_{irr}$.  }
\end{figure}


\newpage

\subsection*{Sequential read-outs}

\begin{figure}[h!]
\includegraphics[width=\columnwidth]{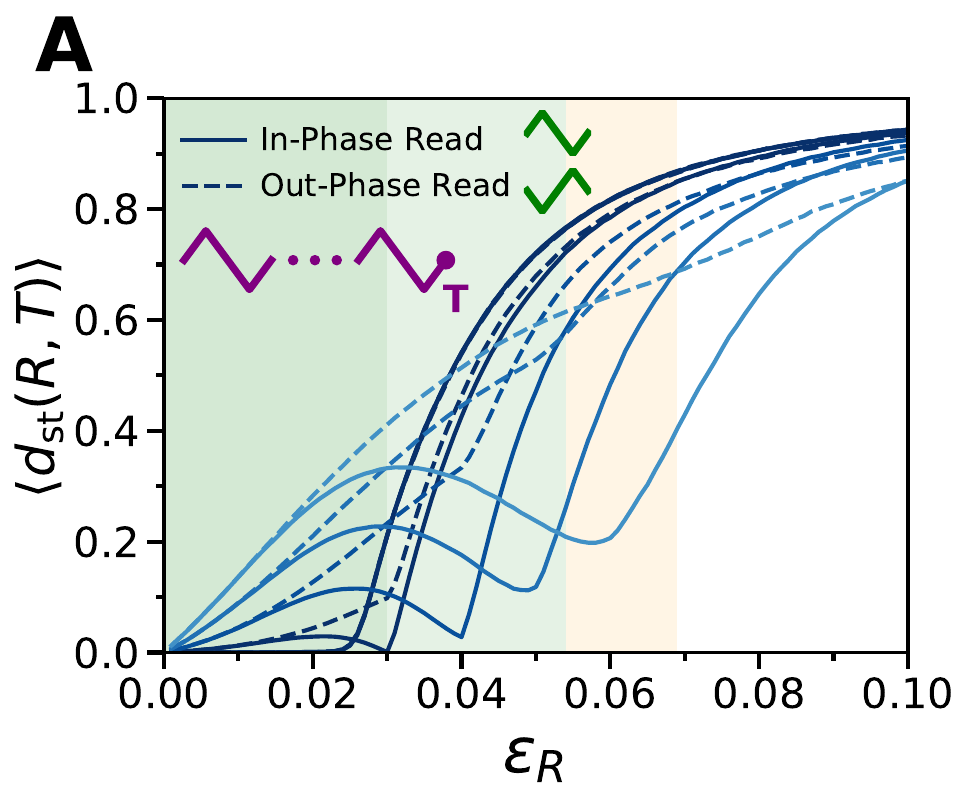}
\includegraphics[width=\columnwidth]{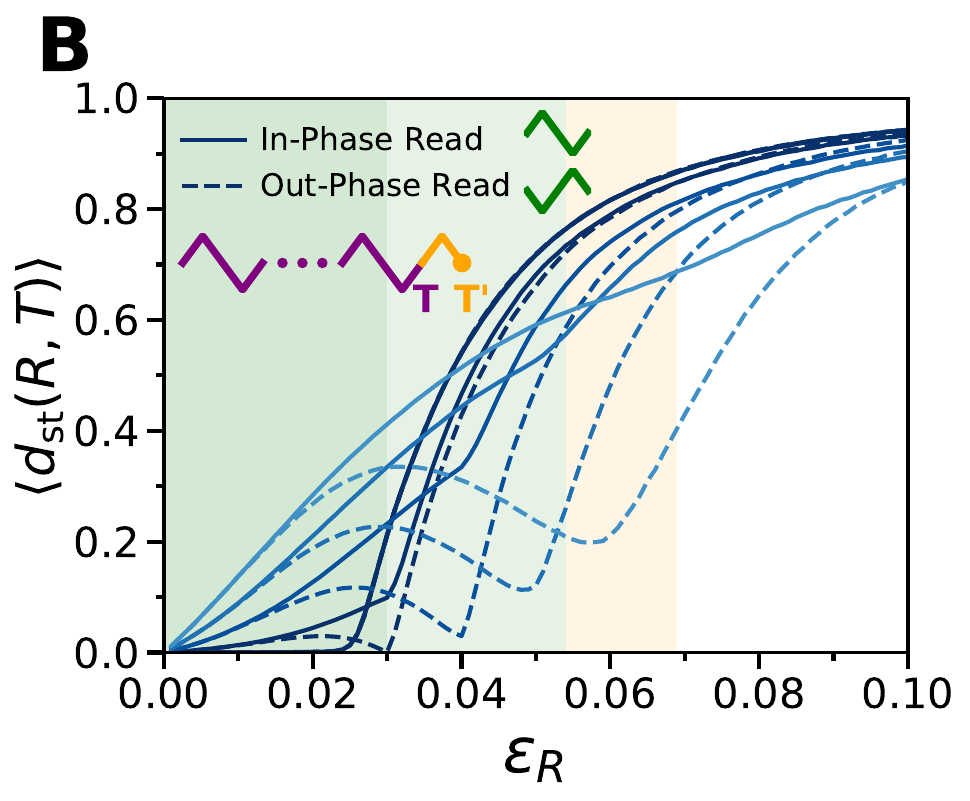}
\caption{In- and out-of-phase sequential read-outs from the trained states $T$ and $T'$.}
\end{figure}

\newpage


\subsection*{Parallel read-outs after adding an out-of-phase training half-cycle}

\begin{figure}[h!]
\centering
\includegraphics[width=\columnwidth]{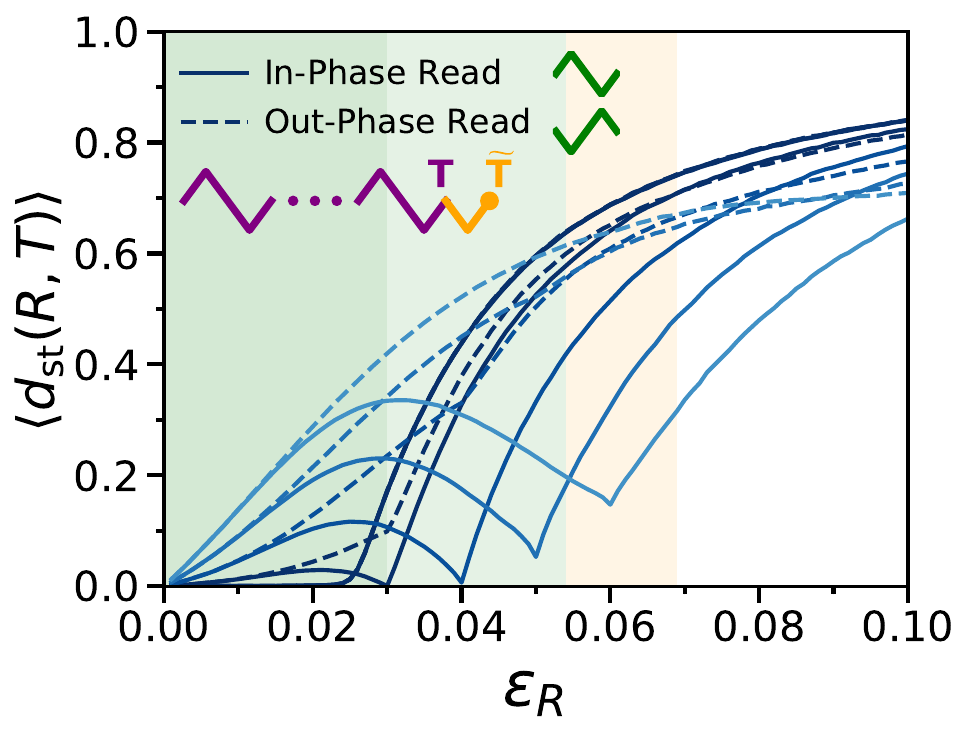}
\caption{Parallel in- and out-phase read-out performed on the state $\widetilde{T}$, obtained by applying an out-of-phase training half-cycle $0 \to -\varepsilon_T \to 0$ to $T$}
\end{figure}

\newpage

\subsection*{Sequential mid-cycle read-out protocol}

\begin{figure}[h!]
\centering
\includegraphics[width=\columnwidth]{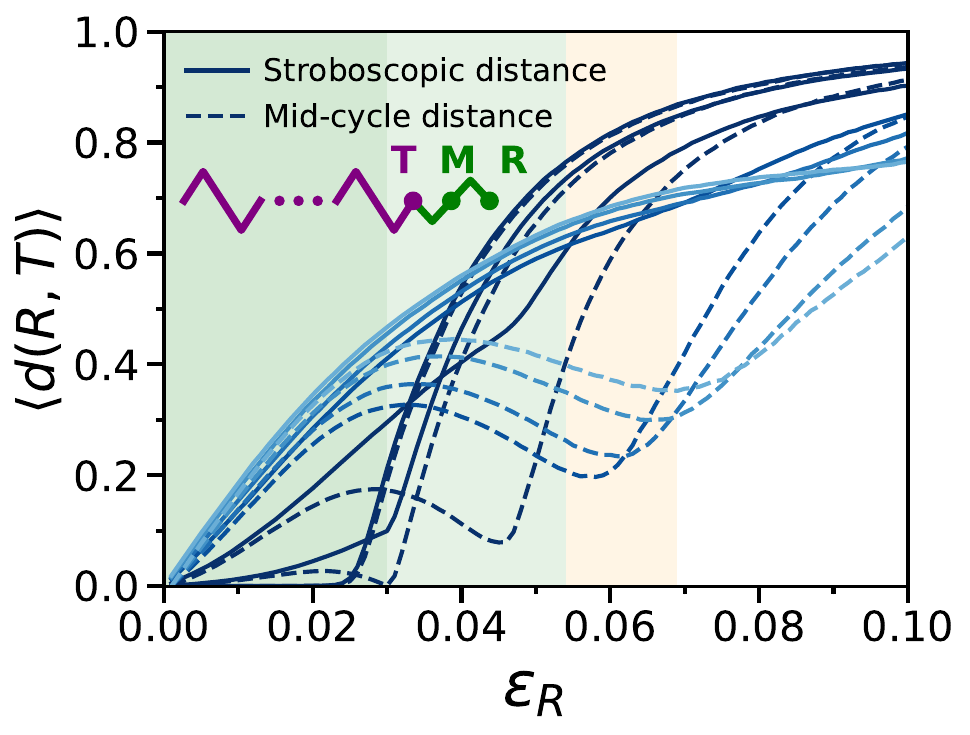}
\caption{Results of the out-of-phase sequential read-out protocol that includes the median state $M$, which is reached at then end of the first half  of the read-out cycles. The stroboscopic distance between $T$ and $M$ is denoted by $d_m$. }
\end{figure}

\end{document}